\documentclass[showpacs,eqsecnum,aps]{revtex4}
\usepackage{amsmath}
\usepackage{amssymb}
\usepackage{amsfonts}
\usepackage{graphics}
\def\beq{\begin{equation}}
\def\eeq{\end{equation}}
\def\beqa{\begin{eqnarray}}
\def\eeqa{\end{eqnarray}}

\def\bLam{\mbox{\boldmath$\Lambda$}}

\def\etab{\mbox{\boldmath$\eta$}}
\def\bLam{\mbox{\boldmath$\Lambda$}}

\def\sumint{\sum\mspace{-25mu}\int}

\def\p{{\sf p}}

\def\x{{\sf x}}

\def\y{{\sf y}}

\def\a{{\sf a}}

\def\X{{\sf X}}

\def\sumint{\sum\mspace{-25mu}\int}

\usepackage[dvips]{graphicx,color}

\begin{document}

\title{A Euclidean formulation of relativistic quantum mechanics}
\author{P. Kopp}
\affiliation{
Department of Physics and Astronomy, The University of Iowa, Iowa City, IA
52242}

\author{W. N. Polyzou}
\affiliation{
Department of Physics and Astronomy, The University of Iowa, Iowa City, IA
52242}

\vspace{10mm}
\date{\today}

\begin{abstract}

  In this paper we discuss a formulation of relativistic quantum
  mechanics that uses Euclidean Green functions or generating
  functionals as input.  This formalism has a close relation to
  quantum field theory, but as a theory of linear operators on a
  Hilbert space, it has many of the advantages of quantum mechanics.
  One interesting feature of this approach is that matrix elements of
  operators in normalizable states on the physical Hilbert space can
  be calculated directly using the Euclidean Green functions without
  performing an analytic continuation.  The formalism is summarized in
  this paper.  We discuss the motivation, advantages and difficulties
  in using this formalism.  We discuss how to compute bound states,
  scattering cross sections, and finite Poincar\'e transformations
  without using analytic continuation.  A toy model is used to
  demonstrate how matrix elements of $e^{-\beta H}$ in normalizable
  states can be used to construct-sharp momentum transition matrix
  elements.
 
\end{abstract}

\vspace{10mm}


\maketitle





\section{Motivation}
 
In this paper we investigate a framework for constructing relativistic
quantum mechanical models of few-degree-of-freedom systems that are
inspired by an underlying quantum field theory.  Our interest
is physics at the few GeV energy scale.  Poincar\'e invariance is an
important symmetry at this scale since the energies are comparable
to the mass scale of hadrons.  Most models of systems at these energies
are motivated by quantum field theory, but their connection with
quantum mechanical models of a finite number of degrees of freedom is
not straightforward.  The advantage of a quantum mechanical model of
a finite number of degrees of freedom is that the theory is linear and
it can be in-principle solved, even for strongly interacting systems,
with mathematically controlled errors.  Bound systems of particles
present no special problems; they are just point-spectrum eigenstates
of the mass operator (rest energy).  The construction of a unitary
multichannel $S$ operator can be performed using the same methods that
are used in non-relativistic models.

One approach that has been successfully used to formulate realistic
Poincar\'e invariant quantum mechanical models of few-hadron systems
is Poincar\'e invariant quantum mechanics\cite{Keister:1991sb}.  In
this approach a dynamical unitary representation of the Poincar\'e
group is constructed on a few-particle Hilbert space.  This approach
has been successfully applied to treat a number of few-hadron or few
quark problems.  The virtue of this approach is that it is possible
construct quantum mechanical models with a finite number of degrees of
freedom that have a unitary representation of the Poincar\'e group,
satisfy a spectral condition, and for fixed numbers of particles
satisfy cluster separability.  These are essentially all of the axioms
of quantum field theory, except microscopic locality, which cannot be
tested experimentally and requires an infinite number of degrees of
freedom.  One of the disadvantages of this approach is that the
models do not have a straightforward relation to a Lagrangian field
theory.  This makes it difficult to use a field theory like QCD to
systematically improve or constrain the models.

Because of the difficulties discussed in the previous paragraph it is
desirable to explore alternate formulations of relativistic quantum
mechanics that have a more direct relation to Lagrangian field theory
while preserving the structure of the underlying quantum theory.  The
alternative that we pursue in this work is to formulate models that
are motivated by the standard reconstruction of a quantum theory from
the field theory.  The Euclidean formulation of quantum field theory
is convenient for this purpose because (1) is has a direct relation to
the action or Lagrangian through either formal path integrals or the
Dyson expansion and (2) it is possible to use the Euclidean Green
functions or generating functional to directly construct the physical
Hilbert space using the Euclidean reconstruction theorem.  In this
paper we argue that using this framework it is possible to compute all
interesting quantum mechanical observables without explicit analytic
continuation.

The proposed models are constructed by restricting the number of
degrees of freedom.  In passing to a few-body model all of the axioms
of field theory cannot be preserved.  One of the attractive features
of the Euclidean reconstruction theorem is that locality is an
independent requirement that can be relaxed without violating the
other axioms\cite{Osterwalder:1973dx}.  A feature of this approach is
that the dynamics is introduced directly by the model Euclidean Green
functions or generating functional, rather than in a model Hamiltonian
or Lagrangian.  One of the challenges of this approach is to find a
robust framework for modeling Euclidean Green functions or generating
functionals with the required properties.  An advantage is that
standard Euclidean field theoretic methods can be used to formulate
models.

The construction of the physical Hilbert space of the field theory
from a Euclidean generating functional is discussed in the next
section.  The generating functional for a scalar field is used to
illustrate the basic construction.  The construction of the Poincar\'e
Lie algebra is discussed in the third section.  The resulting
generators are self-adjoint operators on the physical Hilbert space.
The construction of one-particle states and the computation of finite
Poincar\'e transformations on these states is discussed in section
four.  The one-particle states are used to construct scattering states
in section five.  The computation of finite Poincar\'e transformations
on the scattering states are also discussed in this section.
Everything discussed in sections 2-5 uses only Euclidean generating
functionals and Euclidean test functions.  Section six discuss how the
results of sections 2-5 are expressed in terms of Euclidean Green
functions.  This is more practical for formulating models.  Models of
finite number of degrees of freedom are discussed in section seven by
considering the structure of models of relativistic nucleon-nucleon
scattering.  Numerical tests of the proposed method to compute
scattering observables from matrix elements of $e^{-\beta H}$ in
normalizable states are given in section eight using an exactly
solvable non-relativistic model.  The results suggest that these
methods can be used to compute cross sections at the few GeV energy
scale.  Our conclusions and the outlook for this approach 
are discussed in the last section.

\section{Quantum mechanics - Hilbert space representation} 

To illustrate the construction of the model quantum theory we assume
that we are given an Euclidean generating functional $Z[f]$ associated
with a scalar field.  For the purpose of illustration we assume that
$Z[f]$ has all of the properties that are expected of the generating
functional of a scalar field theory.  These properties include
Euclidean invariance, reflection positivity, reality, and cluster
separability.

The generating functional relates the Lagrangian of the field theory
to the Hilbert space formulation of the theory.  In Lagrangian field
theory the generating functional is formally the functional Fourier transform
of an Euclidean path-integral measure with action $A[\phi]$
\beq
Z[f] = \frac{\int D[\phi] e^{-A[\phi] + i \phi(f)}}{ 
\int D[\phi] e^{-A[\phi]}} = 
\sum_n \frac{(i)^n}{ n!} S_n \underbrace{(f,\cdots ,f)}_{\mbox{$n$ times}} =
\exp (\sum_n \frac{(i)^n }{ n!} S^c_n \underbrace{(f,\cdots ,f)}_{\mbox{$n$ times}})
\label{b.1}
\eeq
where $f= f(\tau, \mathbf{x})$ is a suitable Euclidean test function
and $S_n$ is the $n$-point Euclidean Green function, and $S_n^c$ is
the connected $n$-point Euclidean Green function.  In this approach
the generating functional $Z[f]$ is the dynamical input that replaces
the Lagrangian or Hamiltonian.  A connection with an underlying
Lagrangian or path integral is not required, however this connection
is an important source of phenomenology, which is the primary
motivation for developing this formalism.  

In what follows Euclidean space-time coordinates are denoted by
$\x=(\tau,\mathbf{x})$.  Euclidean invariance of the generating functional 
means that
\beq
Z[f_{E,a}]=Z[f].
\label{b.2}
\eeq
for 
\beq
f_{E,a}(\x) = f\left(E^{-1}(\x-\a)\right),
\label{b.3}
\eeq
where $E$ is an $O(4)$ rotation and $\a$ is a Euclidean space-time 
displacement.

The construction of the physical Hilbert space from Euclidean Green
functions was given by Osterwalder and Schrader
\cite{Osterwalder:1973dx}\cite{Osterwalder:1974tc}.  A simpler
construction in terms of the generating functional was given by
Fr\"ohlich\cite{Frohlich:1974}.  We use Fr\"ohlich's approach to
illustrate the construction using the generating functional for a
``scalar field''.  To construct the physical Hilbert space Osterwalter
and Schrader select a Euclidean time axis and restrict the space of
test functions, $f(\tau,\mathbf{x})$, to real-valued Schwartz
functions of four Euclidean variables that vanish for negative
Euclidean times;
\beq
{\cal S}_+ = \{ f(\tau, \mathbf{x}) \in {\cal S} \vert f(\tau,
\mathbf{x}) = 0 \quad \tau < 0 \}.
\label{b.4}
\eeq
We call elements of ${\cal S}_+$ positive-time test functions.  In
what follows all test functions will be assumed to be positive-time
test functions unless stated otherwise.

Osterwalter and Schrader introduce the Euclidean time-reflection
operator $\Theta$ defined by
\beq
\Theta f(\tau, \mathbf{x}) = f(-\tau, \mathbf{x}).
\label{b.5}
\eeq
The generating functional, $Z[f]$, is reflection positive if
for any finite sequence of real positive-time test functions, $\{ f_1
\cdots f_n\} \in {\cal S}_+ $, the $n\times n$ matrix
\beq
M_{ij} := Z[f_i - \Theta f_j ] \geq 0;
\label{b.6}
\eeq
is non-negative.  This condition is not automatic.  It holds for
generating functionals for free fields and for some lattice
truncations of interacting theories.  In general it is a requirement
on physically acceptable models.  In what follows we simply assume
this condition is satisfied.

In Fr\"ohlich's construction a dense set of normalizable vectors in the 
physical Hilbert space is represented by complex wave-functionals of the form
\beq
B[\phi] = \sum_{j=1}^{n_b} b_j e^{i \phi (f_j) }
\label{b.7}
\eeq
where $b_i$ are complex numbers and $f_j(\tau, \mathbf{x})$ are real
positive-time test functions.  The argument $\phi$ can be thought of
as an abstract integration variable.  This interpretation is motivated
by the path integral representation of the generating functional.

The physical scalar product of two such wave-functionals, $B[\phi]$, and 
\beq
C[\phi] =  \sum_{k=1}^{n_c} c_k e^{i \phi (g_k) }
\label{b.8}
\eeq
is given directly in terms of the generating functional by
\beq
\langle B \vert C \rangle :=
\sum_{j,k}^{n_b,n_c}  b_j^* c_k Z[g_k-\Theta f_j].
\label{b.9}
\eeq
Reflection positivity, (\ref{b.6}),  is equivalent to the statement 
\beq
\langle B \vert B \rangle \geq 0 .
\label{b.10}
\eeq
The physical Hilbert space is obtained by identifying wave functions
whose difference has zero norm and completed by adding convergent
sequences of wave functionals.

The inner product (\ref{b.9}) is the physical quantum mechanical
scalar product, even though the input only involves the Euclidean
generating functional and positive-time Euclidean test functions!  No
analytic continuation is used.  An explicit illustration of this
relationship in given in section six for free scalar particles in
eq. (\ref{f.2a}) and free spin $1/2$ particles in eq. (\ref{f.2b}).

While the computation of the exact generating functional is equivalent
to solving the field theory, models of generating functionals are
easily constructed.  For example, consider the representation of the
generating functional in terms of connected Euclidean Green functions.
If we isolate the contribution to the generating functional from the
connected $n$-point Green function (see \ref{b.1}),
\beq
Z^c_n [f] := e^{\frac{(i^n)}{ n!}S^c_n (f, \cdots ,f)},  
\eeq
then the full generating functional is the product
\beq
Z [f] = \prod_n Z^c_n [f].
\eeq
It follows that the matrix that gives the quantum mechanical scalar product,
(\ref{b.6}),  has the decomposition
\beq
M_{ij} := Z [f_i-\theta f_j ] = \prod_n Z^c_n [f_i - \theta f_j]=
\prod_n M^n_{ij} 
\label{b.10b}
\eeq
which is an infinite Schur product of the matrices $M^n_{ij}$.  A
sufficient condition for positivity of $M_{ij}$ is that each
$M^n_{ij}$ in the Schur product is positive (this is because the Schur
product can be expressed as the restriction of the tensor product of
positive operators to the diagonal subspace).

Thus, one strategy for constructing models is to use the
representation (\ref{b.10b}),  to build models starting with the free
field two-point function.  $Z^c_2 [f]$ is reflection positive if it is
the generating functional for a generalized free field.  Including a
reflection positive $Z^c_4 [f]$ in the Schur product (\ref{b.10b}) 
would give a generating functional for an interacting many-body 
theory, where the
dynamical input is a suitable connected Euclidean 4-point function,
\beq
Z_{model} [f] =  Z^c_2 [f] Z^c_4 [f].
\eeq
Positivity of $M^4_{ij}$ means that that
\beq
Z^c_4 [f_i - \theta f_j] =e^{\frac{i^4 }{ 4!}S_4^c(
f_i - \theta f_j,f_i - \theta f_j,f_i - \theta f_j,f_i - \theta f_j)}
\eeq
is a positive matrix for any finite sequence of positive-time test
functions.  The model can be refined by including additional factors,
$Z^c_n[f]$. 

\section{Relativistic invariance} 

Poincar\'e invariance of a quantum theory implies the existence of a
unitary representation of the Poincar\'e group on the physical Hilbert
space \cite{Wigner:1939cj}.  Equivalently there should be a set of ten
infinitesimal generators of the Poincar\'e group represented by
self-adjoint operators satisfying the Poincar\'e commutation
relations.
 
The relation between the complex Euclidean group and the complex 
Poincar\'e group is relevant for constructing Poincar\'e generators.
To understand the connection between the 4-dimensional Euclidean group 
and the Poincar\'e group consider the matrices
\beq
X=
\left ( 
\begin{array}{cc} 
t-z & x-iy \\
x+iy & t+z 
\end{array} 
\right ) 
\qquad
\X = \left ( 
\begin{array}{cc} 
i \tau-z & x-iy \\
x+iy & i \tau +z 
\end{array} 
\right ) .
\label{c.1}
\eeq
A simple calculation shows that $\mbox{det} (X) = t^2 - \mathbf{x}^2$ and 
$\mbox{det}(\X) = - (\tau^2 + \mathbf{x}^2)$, which are the 
Minkowski and Euclidean invariant (distances)$^2$ respectively.   
Both determinants are preserved under the linear transformations
\beq
X\to X'=AXB^t \qquad \X \to \X' = A\X B^t \qquad \mbox{det} (A) = \mbox{det}(B) =1 
\label{c.2}
\eeq
where $A$ and $B$ are complex matrices with unit determinant.  In
general the pair $(A,B)$ defines a complex Lorentz or complex $O(4)$
transformation.  Both $(A,B)$ and $(-A,-B)$ correspond to the same
linear transformation of the coordinates.  In general, the transformed
coordinates may become complex, but the determinant remains real and
unchanged.  If $B=A^*$ then transformation $X'=AXB^t$ is a real
Lorentz transformation; if $A,B\in SU(2)$ then the transformation
$\X'=A\X B^t$ is a real $O(4)$ transformation.

When $A,B\in SU(2)$ the transformation $X'=AXB^t$ is a six parameter
subgroup of the complex Lorentz group.  While this represents a real
Euclidean transformation on the Euclidean Hilbert space
(without the Euclidean time reversal $\Theta$), it defines a
complex Lorentz transformation on the physical Hilbert space.  It is
possible to extract the ten Poincar\'e generators by considering
infinitesimal forms of these complex Lorentz transformations.

In order to get a unitary representation of the real Poincar\'e group
on the physical Hilbert space the generators must be self-adjoint.  It
turns out that the $\Theta$ in the physical scalar product breaks
Euclidean invariance in just the right way to ensure that all of the
Poincar\'e generators are self-adjoint.

We begin by defining real Euclidean transformations on wave functionals by
\beq
\left (T(E,\a)B\right )[\phi]  = \sum_{j=1}^{n_b} b_j e^{i \phi (f_{j,E,a}) }
\qquad 
f_{j,E,\a} (\x) := f_j\left ( E^{-1}(\x-\a)\right ) .
\label{c.3}
\eeq
These transformations leave the generating functional
invariant, $Z[f]=Z[f_{E,\a}]$.  In general they will not preserve
the positive time constraint.

Before we use these transformations to construct Poincar\'e generators
on the physical Hilbert space, it is useful to note that the wave
functionals can also be considered as multiplication operators.  For
example the operator $B[\phi]$ acting on the wave functional $C[\phi]$
is a wave functional $D[\phi]$ defined by
\beq
D[\phi] := B[\phi]C[\phi]= \sum_{j=1}^{n_b}\sum_{k=1}^{n_c} b_j c_k 
e^{i \phi(f_j+g_k)}.
\label{c.4}
\eeq
This will become important when we formulate scattering asymptotic
conditions.  Note that in the wave functional representation the order
of the Minkowski field operators is determined by the order of the
Euclidean time support of the positive time Schwartz functions.

Next we consider the real Euclidean transformations, $T(E,\a)$, as
complex Poincar\'e transformations on the physical Hilbert space.  It
is useful to treat pure rotations, space translations, Euclidean time
translations, and Euclidean rotations in planes containing the time
axis separately.  The Euclidean time reversal $\Theta$ operator does
not commute with the last two transformations.
 
We define action of the Poincar\'e generators on the wave 
functionals considered as operators by:
\beq
[\mathbf{P}, B[\phi]]  
=-i \frac{\partial }{ \partial \mathbf{a}} [T(I,(0,a)),B][\phi]_{\vert_{\mathbf{a}=0}}  
:= 
-i \sum_{j=1}^{n_b} b_j 
\frac{\partial}{\partial \mathbf{a}} 
e^{i \phi (f_{j,I,(0,\mathbf{a})})}_{\vert_{\mathbf{a}=0}}{},
\label{c.5}
\eeq
\beq
[\mathbf{J}\cdot \hat{\mathbf{n}},  B[\phi]]  
=-i \frac{\partial}{\partial \xi} (T(R(\hat{\mathbf{n}},\xi),0)B)[\phi]_{\vert_{\xi=0}}   := 
-i \sum_{j=1}^{n_b} b_j 
\frac{\partial }{\partial \xi} 
e^{i \phi (f_{j,(R(\hat{\mathbf{n}},\xi),0)})}_{\vert_{\xi=0}}{}.
\label{c.6}
\eeq
where $R(\hat{\mathbf{n}},\xi)$ is an ordinary rotation about the 
$\hat{\mathbf{n}}$ axis by an angle $\xi$.
For the Hamiltonian we require that $\beta>0$ in $T(I,(\beta,a))$ to 
preserve the positive-time support condition: 
\beq
[H ,B[\phi]]  
=- \frac{\partial }{ \partial \beta} (T(I,(\beta,0))B)[\phi]_{\vert_{\beta=0}}  
:= 
-\sum_{j=1}^{n_b} b_j 
\frac {\partial }{ \partial \beta} 
e^{i \phi (f_{j,I,(\beta,0)})}_{\vert_{\beta=0}}{}.
\label{c.7}
\eeq
For the boost generators we first restrict the support of the 
test functions $f_j$ in the wave functionals to a cone symmetric
about the positive Euclidean time axis that makes an 
angle $0 \leq \chi< \pi/2$ with the Euclidean time axis 
\beq
{\cal S}_{\chi,+} := \{ f \in {\cal S}_+ \, \vert \,
f (\tau,\mathbf{x}) =0 \quad  
\tan^{-1} ( \frac{\tau}{\vert \mathbf{x} \vert}) \geq \chi  \}{}.
\label{c.8}
\eeq
This ensures that the support condition is preserved for
sufficiently small rotations. 
On these wave functionals we consider the Euclidean
rotation $T(R_e(\hat{\mathbf{n}},\xi),0)$ in the 
$\hat{\mathbf{n}}-\tau$ plane through angle $\xi < \pi/2 - \chi$: 
\beq
f_{j,\xi,\hat{\mathbf{n}}} (\tau,\mathbf{x} ) := 
f_j (\tau',\mathbf{x}' ) \qquad f_f\in {\cal S}_{\chi,+},
\label{c.9}
\eeq
with 
\beq
\tau'=\tau \cos(\xi)-x_{\hat{\mathbf{n}}} \sin (\xi)
\qquad x_{\hat{\mathbf{n}}}' = x_{\hat{\mathbf{n}}} 
\cos (\xi) + \tau \sin (\xi) .
\label{c.10}
\eeq
The restrictions on the parameters $\xi$ and $\chi$ ensure that 
initial and final vectors are in the physical Hilbert space.
On these vectors the rotationless boost generator is defined by
\beq
[\mathbf{K}\cdot \hat{\mathbf{n}} , B[\phi]]  
=- {\partial \over \partial \xi} (T(R_e(\hat{\mathbf{n}},\xi),0)B)[\phi]_{\vert_{\xi=0}}   := 
-\sum_{j=1}^{n_b} b_j 
 {\partial \over \partial \xi} 
e^{i \phi (f_{j,(R_e(\hat{\mathbf{n}},\xi),0)})}_{\vert_{\xi=0}}
\label{c.11}
\eeq
where $R_e(\hat{\mathbf{n}},\xi)$ is the Euclidean space-time
rotation (\ref{c.10}).
Note the absence of the $i$ in the expressions for $H$ and $\mathbf{K}$.
This is compensated for by the $\Theta$ that appears in the 
physical scalar product.

Direct calculations show that the ten operators
$H,\mathbf{P},\mathbf{J},\mathbf{K}$ satisfy the Poincar\'e
commutation relations and are formally Hermitian on the physical
Hilbert space.  Self-adjointness of $H,\mathbf{P},\mathbf{J}$ follows
because these operators are generators of either one parameter unitary
groups or a contractive Hermitian semigroup.    
The contractive nature of time Euclidean time
evolution is proved using reflection positivity and positivity
properties of the generating functional\cite{glimm:1981}.  This also 
ensures that Hamiltonian satisfies the spectral condition:
\beq
H \geq 0 .
\label{c.12}
\eeq
Self-adjointness of $\mathbf{K}$
can be established by verifying that it is the generator of
a local symmetric
semigroup \cite{Klein:1981}\cite{Klein:1983}\cite{Frohlich:1983kp}.

Matrix elements of the Poincar\'e generators in normalizable states
can be expressed directly in terms of the generating functional
\beq
\langle B \vert H \vert C \rangle = 
-{\partial \over \partial \beta}
\left ( \sum_{j=1}^{N_b} \sum_{k=1}^{N_c} 
b^*_j c_k Z[ g_{k,I,(\beta,0)} - \Theta f_j  ] \right )_{\beta=0} ,
\label{c.13}
\eeq
\beq
\langle B \vert \mathbf{P} \vert C \rangle =
-i{\partial \over \partial \mathbf{a}}
\left (
\sum_{j=1}^{n_b} \sum_{k=1}^{n_c} 
b^*_j c_k Z[ g_{k,I,(0,\mathbf{a})} - \Theta f_j ] 
\right )_{\mathbf{a}=0} ,
\label{c.14}
\eeq
\beq
\langle B \vert \hat{\mathbf{n}} \cdot \mathbf{J} \vert C \rangle
=
-i{\partial \over \partial \xi}
\left (
\sum_{j=1}^{b_b} \sum_{k=1}^{n_c} 
c^*_j d_k Z[g_{k,R(\hat{\mathbf{n}},\xi),0} - \Theta f_j  ] 
\right )_{\xi=0} ,
\label{c.15}
\eeq
\beq
\langle B \vert \hat{\mathbf{n}} \cdot \mathbf{K} \vert C \rangle
=
-{\partial \over \partial \xi}
\left (
\sum_{j=1}^{b_b} \sum_{k=1}^{n_c} 
c^*_j d_k Z[g_{k,R_e(\hat{\mathbf{n}},\xi),0} - \Theta f_j  ] 
\right )_{\xi=0} .
\label{c.16}
\eeq

The formal Hermiticity of the generators defined above can be deduced from
these expressions.  For example
\[
\langle B \vert H^{\dagger}  \vert C \rangle = 
\langle C \vert H  \vert B \rangle^* = 
\]
\[
-{\partial \over \partial \beta}
\left ( \sum_{j=1}^{N_b} \sum_{k=1}^{N_c} 
b_j c^*_k Z[- f_{k,I,(\beta,0)} + \Theta g_j  ] \right )_{\beta=0} =
\]
\[
-{\partial \over \partial \beta}
\left ( \sum_{j=1}^{N_b} \sum_{k=1}^{N_c} 
b_j c^*_k Z[- f_{k} + (\Theta g)_{j,I,(-\beta,0)}] \right )_{\beta=0} =
\]
\beq
\left ( \sum_{j=1}^{N_b} \sum_{k=1}^{N_c} 
b_j c^*_k Z[\Theta g_{j,I,(\beta,0)} - f_{k}  ] \right )_{\beta=0} =
\langle B \vert H  \vert C \rangle 
\label{c.17a}
\eeq
where we have used reality, $Z^*[f]=Z[-f]$, Euclidean invariance, 
and properties of $\Theta$.   Hermiticity of the rotationless 
boost generators follows using the same argument.  The 
Euclidean time-reversal operator, $\Theta$,  plays the role of the missing 
factor of $i$ when integrating by parts.

The commutation relations can be verified by explicit computation, 
however they also follow as a direct 
consequence of the relation between complex $O(4)$ and the 
complex Lorentz group. 

Matrix elements of $e^{-\beta H}$ can also be directly computed in terms of the
generating functional:
\beq
\langle B \vert e^{-\beta H} \vert C \rangle = 
\sum_{j=1}^{N_b} \sum_{k=1}^{N_c} 
b^*_j c_k Z[g_{k,I (\beta,0)} - \Theta f_j   ] .
\label{c.17}
\eeq
These matrix elements are needed to compute scattering cross sections
and only involve elementary quadratures. 

Matrix elements of the mass Casimir operator can be expressed in terms of the 
Poincar\'e generators 
\beq
M^2 = H^2 -\mathbf{P}^2 
\label{c.18}
\eeq
\beq
\langle B \vert M^2 \vert C \rangle := \left ( {\partial^2 \over \partial \beta^2} +
{\partial^2 \over \partial \mathbf{a}^2} \right )
\sum_{j=1}^{N_b} \sum_{k=1}^{N_c} 
b^*_j c_k Z[g_{k,I (\beta,\mathbf{a})} - \Theta f_j   ]_{\vert_{\beta=\mathbf{a}=0}} .
\label{c.19}
\eeq 
Finally we note that the real Euclidean transformations, $T(E,\a)$ can 
be expressed in terms of the Poincar\'e generators on the physical 
Hilbert space by
\beq
T(I,(\beta ,\mathbf{a})) = e^{-\beta H + i \mathbf{a} \cdot \mathbf{P}}
\qquad
T(R(\hat{\mathbf{n}},\psi ),0) = e^{i \mathbf{J} \cdot \hat{\mathbf{n}} \psi }
\qquad 
T(R_e(\hat{\mathbf{n}},\psi ),0) = e^{ \mathbf{K} \cdot \hat{\mathbf{n}}\psi} .  
\label{c.20}
\eeq 
Thus Euclidean time evolution and rotations in Euclidean space-time 
planes look like imaginary time evolution and Lorentz transformation with
imaginary rapidities.

The operators defined in (\ref{c.5},\ref{c.6},\ref{c.7} and
\ref{c.11}) are self-adjoint operators on the physical Hilbert space
that satisfy the Poincar\'e commutation relations.  Formally they can
be exponentiated to give a unitary representation of the Poincar\'e
group on the physical Hilbert space but, as we will see in the next
two sections, this exponentiation is never needed.

The expressions for the matrix elements of all of the Poincar\'e
generators in normalizable states (\ref{c.13}-\ref{c.16}) are directly
expressed in terms of the Euclidean generating functional and
Euclidean test functions.  Analytic continuation is not used.

\section{Particles} 

Given the dense set of wave functionals of the form (\ref{b.7}) and
the physical scalar product (\ref{b.9}) the Gram-Schmidt method can be
formally used to construct a complete orthonormal set of wave
functionals $B_n [\phi]$:
\beq
\langle B_n \vert B_m \rangle =\delta_{mn} .
\label{d.1}
\eeq
Since the orthonormal wave functionals are complete, 
normalizable one-particle states are 
linear combinations of these orthonormal wave functionals
with square summable coefficients: 
\beq
\Psi_{\lambda}  [\phi] = \sum_n b_n B_n [\phi] 
\qquad 
\sum_n \vert b_n\vert^2 < \infty 
\label{d.2}
\eeq
that are eigenstates of the mass square Casimir operator (\ref{c.18}) of the 
Poincar\'e group with eigenvalue $\lambda^2$ in the point spectrum:
\beq
\sum_n \langle B_m \vert M^2\vert B_n \rangle b_n =
\sum_n \langle B_m \vert (H^2-\mathbf{P}^2)\vert B_n \rangle b_n = \lambda^2
\delta_{mn} b_n . 
\label{d.3}
\eeq

These normalizable states are infinitely degenerate because there is an 
associated wave packet in the particle's momentum and spin.
For suitable wave packets these normalizable eigenstates 
can be decomposed into simultaneous eigenstates of mass and linear 
momentum using translations and Fourier transforms:
\beq
\Psi_{\lambda, \mathbf{p}}[\phi]=
\int {d\mathbf{a} \over (2 \pi )^{3/2}} e^{-i \mathbf{p} \cdot \mathbf{a}}
\left ( T(0, \mathbf {a})\Psi_\lambda 
\right ) [\phi] .
\label{d.4}
\eeq
These wave functionals can be given a plane-wave normalization  
\beq
\langle 
\Psi_{\lambda, \mathbf{p}'} \vert \Psi_{\lambda, \mathbf{p}}
\rangle =
\delta (\mathbf{p}'-\mathbf{p}) .
\label{d.15}
\eeq

The simultaneous eigenstates of mass and linear
momentum can be further decomposed into eigenstates of spin, and
$z$-component of spin using
\beq
\Psi_{\lambda, j, \mathbf{p},\mu }[\phi]=
\sum_{\nu=-j}^j \int_{S(U2)}  dR
\left (T(R,0) \Psi_{\lambda, R^{-1} \mathbf{p}} \right )   
[\phi]D^{j*}_{\mu \nu}(R) 
\label{d.6}
\eeq
where the integral is over $SU(2)$ and $dR$ is the $SU(2)$ Haar
measure.  This projection gives the canonical spin.  This is the spin
measured in the rest frame when the particle is transformed to the
rest frame by a rotationless Lorentz transformation.  Different
projections can be used to get states of different helicities or
light-front spins.  The integral in equation (\ref{d.6}) vanishes 
if there are no states of mass $\lambda$ and spin $j$.

The normalization of the states can be chosen so 
\beq
\langle \Psi_{\lambda, j', \mathbf{p}',\mu' }\vert 
\Psi_{\lambda, j, \mathbf{p},\mu }\rangle
=
\delta (\mathbf{p}'-\mathbf{p})\delta_{j'j}\delta_{\mu'\mu} .
\label{d.7}
\eeq
The state  $\vert \Psi_{\lambda, j, \mathbf{p},\mu } \rangle$ is a single particle
state if $\lambda$ is in the discrete spectrum of $M$.

Since we started from a linear combination of wave functionals, the 
single-particle state is formally represented by a single-particle 
wave functional 
\beq
\Psi_{\lambda, j', \mathbf{p}',\mu' } [\phi] .
\label{d.8}
\eeq

In general it is not trivial to compute finite Poincar\'e transformations
in terms of the generators, however if the one-particle state is a
non-degenerate state (i.e. the theory has no other particles with the
same mass and spin) then this state necessarily transforms irreducibly
with respect to the dynamical unitary representation of the Poincar\'e
group.  It follows that
\[
\langle B \vert U(\Lambda ,a)\vert  \Psi_{\lambda ,j, \mathbf{p}, \mu}  \rangle =
\]
\beq
\langle B \vert \psi_{\lambda  ,\mathbf{p}', \mu'}  \rangle
\sqrt{{\omega_{\lambda} (\mathbf{p}') \over 
\omega_{\lambda} (\mathbf{p})}} 
e^{-i \omega_{\lambda}(\mathbf{p}')a^0 + i \mathbf{p}'\cdot \mathbf{a}} 
D^j_{\mu' \mu} [\Lambda_c^{-1}({\mathbf{p}'\over \lambda})\Lambda   
\Lambda_c({\mathbf{p}\over \lambda})]
\label{d.9}
\eeq
where
\beq
(\mathbf{p}')^j = \Lambda^j{}_0 \omega_{\lambda}(\mathbf{p}) +
\Lambda^j{}_{k} \mathbf{p}^k
\qquad 
\omega_{\lambda}(\mathbf{p}) = \sqrt{\lambda^2 + \mathbf{p}^2} 
\label{d.10}
\eeq
and $\Lambda_c({\mathbf{p}\over \lambda})]$ is a rotationless Lorentz
boost that transforms a frame where a particle of mass $\lambda$ is at 
rest to one where it has linear momentum $\mathbf{p}$.

Normalizable single-particle states have the form
\beq
\Psi_{\lambda, j,f }[\phi]  = \int d\mathbf{p} \sum_{\mu=-j}^j f(\mu,\mathbf{p})  
\Psi_{\lambda, j', \mathbf{p}',\mu' } [\phi] 
\label{d.11}
\eeq
\beq
\langle \Psi_{\lambda, j,g } \vert \Psi_{\lambda, j,f } \rangle = 
\int d\mathbf{p} \sum_{\mu=-j}^j 
g^*(\mu,\mathbf{p})
f(\mu,\mathbf{p}).
\label{d.12}
\eeq
In this formalism there is no distinction between elementary particles
and bound states.  They describe particles because they have discrete
mass eigenvalues. 

What is interesting is that it is possible to construct the Hilbert
space, Poincar\'e generators, find single-particle states and perform
finite Poincar\'e transformations on single-particle states using only
the Euclidean generating functional, and positive-time test functions
without performing any analytic continuation.

\section{Scattering theory}

In a quantum theory scattering states are solutions of the
Schr\"odinger equation that evolve into asymptotically separated
non-interacting single-particle states or bound states.  In quantum
field theory there is no free dynamics on the physical Hilbert space,
so with the exception of one-particle states, there are no states
of non-interacting particles on the physical Hilbert space.  However,
because of cluster properties, there are states that look like states
of asymptotically separated particles.  These states evolve like
systems of free particles until the particles get close enough to
interact.

One natural framework to formulate scattering asymptotic conditions
that is applicable in both quantum mechanics and quantum field theory
is the two-Hilbert space formulation of scattering \cite{Coester:1965zz}.
In this framework a separate many-particle Hilbert space of
non-interacting particles is introduced.  This space is used to label
the states of the asymptotically stable particles.  There is a mapping
from this asymptotic space to the physical Hilbert space that adds the
correct description of the internal structure of the particles on the
physical Hilbert space when the particles are asymptotically
separated.

In the asymptotic Hilbert space composite particles are treated like
elementary particles with a given mass, spin, and momentum
distribution.  The internal structure of the composite particle is
contained in the mapping to the physical Hilbert space.  In field
theories all particles have internal structure due to their
self-interactions.  In the non-relativistic case the mapping from the
asymptotic Hilbert space to the physical Hilbert space has the form
\beq
\prod_i \vert (\lambda_i,j_i) \mathbf{p}_i, \mu_i \rangle 
\label{e.1}
\eeq
where the product is a symmetrized tensor product of possibly
composite particle states with given momentum and spin.  Normalizable
states in the physical Hilbert space are obtained when this mapping is
integrated over square integrable functions of the momenta and
magnetic quantum numbers of each asymptotically stable particle.  In
this way composite particles are treated as elementary particles
with mass $\lambda_i$ and spin $j_i$ in the asymptotic Hilbert space
while the mapping adds the composite structure of the asymptotically
separated bound states in the physical Hilbert space.
 
In quantum field theory there are two standard approaches to
scattering.  The most common is the LSZ treatment of scattering, which
formulates the scattering asymptotic conditions using interpolating
fields that create states from the vacuum that have the same quantum
numbers as single particle states.  It has the advantage that the
asymptotic conditions can be formulated without solving the one-body
problem.  The price for this advantage is that weak limits must be
used to calculate scattering matrix elements.  The second approach is
Haag-Ruelle scattering \cite{Haag:1958vt}\cite{Ruelle:1962} which uses
non-local fields that create only one-particle states from the vacuum.
Haag-Ruelle scattering is the natural generalization of standard
quantum mechanical scattering in the field theory setting and it has a
natural two-Hilbert space
formulation\cite{simon}\cite{baumgartl:1983}.  Haag-Ruelle scattering
states are defined by norm limits, just like in the non-relativistic
case.  Haag-Ruelle scattering is not commonly used in applications
because it requires the solution of the one-body problem on the
physical Hilbert space as input.  In this work the one-body solutions
discussed in the previous section are used to formulate the
Haag-Ruelle asymptotic conditions.

We begin with a heuristic summary of the two-Hilbert space formulation
of Haag-Ruelle scattering in Minkowski field theory.  For simplicity
we consider a scalar field theory with a single-particle state of mass
$\lambda$.  To construct the mapping from the asymptotic Hilbert space
to the physical Hilbert space the Fourier transform of the field,
$\tilde{\phi}(p)$, is multiplied by a smooth function $\rho (p^2)$ of
the square of the Minkowski four momentum that is one when $p^2 = -
\lambda^2$ (the mass of the asymptotic particle) and identically
vanishes when $-p^2$ is in the rest of spectrum of $M^2$.

The product of the Fourier transform of the field and the test
function, $\tilde{\phi}_{\rho} (p) := \rho(p^2) \tilde{\phi}(p)$, is
then Fourier transformed back to configuration space.  The resulting
field $\phi_{\rho} (x)$ is covariant, but no longer local.  It has the
property that when it is applied to the physical vacuum it creates
only the single-particle eigenstate of the mass operator with mass
$\lambda$.  Because of the multiplication by $\rho (p^2)$,
$\phi_{\rho} (x)$ is a well-behaved operator-valued function of time
when it is smeared over a test function in three space variables.

The part of $\phi_{\rho} (x)$ that asymptotically looks like a
creation operator is extracted by taking the linear combination of
$\phi$ and $\dot{\phi}$ below:
\beq
A(f,t):= -i \int \phi_{\rho} (x)  \stackrel{\leftrightarrow}{\partial_0}
f (x) d\mathbf{x} 
\label{e.2}
\eeq
where $f(x)$ is a smooth positive-energy solution of the Klein-Gordon
equation. Here smooth means that the Fourier transform of the $t=0$
solution is a smooth function with compact support in the three momentum.
Haag and Ruelle show that the $N$-particle scattering 
states in the physical Hilbert space exist and are given by the strong limits 
\beq
\vert \Psi_{\pm} (f_1, \cdots f_N) \rangle =
\lim_{t \to \pm \infty}   
A(f_N,t) \cdots A(f_1,t) \vert 0 \rangle .
\label{e.3}
\eeq

Next we express the limit (\ref{e.3}) in a two Hilbert space framework
that can also be used in the wave functional representation of the
physical Hilbert space.  First we write $A(f,t)$ defined in
(\ref{e.2}) by expressing the time derivative of the field using the
commutator with the Hamiltonian and the time derivative of the
Klein-Gordon solution by an energy factor:
\[
A(f,t) \vert 0 \rangle = 
\]
\beq
-{i \over (2 \pi)^{3/2}} \int d\mathbf{x}\, d\mathbf{p} e^{i H t} 
\left ( 
i [H, \phi_{\rho} (0,\mathbf{x})] -i \omega_{\lambda}(\mathbf{p}) 
\phi_{\rho} (0,\mathbf{x}) 
\right ) e^{-i Ht} \vert 0 \rangle 
e^{-i \omega_\lambda (\mathbf{p})t + i \mathbf{p} \cdot \mathbf{x} }
\tilde{f}(\mathbf{p})
\label{e.4}
\eeq
where $\tilde{f}(\mathbf{p})$ is a test function in the three momentum.
Integrating over $\mathbf{x}$ gives a partial Fourier transform of 
the field so (\ref{e.2}) becomes 
\beq
A(f,t) \vert 0 \rangle = 
e^{i H t} \int  
\left ( 
[H, \phi_{\rho} (0,\mathbf{p})] - \omega_{\lambda}(\mathbf{p}) 
\phi_{\rho}(0,\mathbf{p}) 
\right ) \vert 0 \rangle 
\tilde{f}(\mathbf{p}) e^{-i \omega_\lambda (\mathbf{p})t}  
d\mathbf{p} .
\label{e.5}
\eeq
The time-dependence is in the quantities $e^{iHt}$ and 
$e^{-i \omega_\lambda (\mathbf{p})t}$, where the second factor
gives the time dependence of the positive-energy solution 
the Klein-Gordan equation.  This is expressed as an operator that 
acts on the wave packet $\tilde{f}(\mathbf{p})$ of a free particle 
of mass $\lambda$.   It follows that 
(\ref{e.2}) can be interpreted as a mapping from a dense subset of 
the Hilbert space of square integrable functions, $f(\mathbf{p})$, to the 
physical Hilbert space 
\beq
A(f,t) \vert 0 \rangle = e^{iHt} A_1 e^{-i H_0t} \vert f \rangle
\label{e.6}
\eeq 
where $H_0=\omega_{\lambda}(\mathbf{p})$ is the energy of the 
asymptotic particle and 
\beq
A_1(\mathbf{p}):= \left ( 
[H, \phi_{\rho} (0,\mathbf{p})] - \omega_{\lambda}(\mathbf{p}) 
\phi_\rho (0,\mathbf{p}) 
\right ).
\label{e.7}
\eeq
By repeating this analysis $N$-times  
the products that appear in the Haag-Ruelle formula, 
\beq
\prod A(f_1,t) \cdots A(f_N ,t) \vert 0 \rangle, 
\label{e.8}
\eeq
can be expressed as mappings from an $N$-particle subspace of 
the Fock space of non-interacting particles of mass $\lambda$ to the 
physical Hilbert space:   
\beq
\prod A(f_1,t) \cdots A(f_N ,t) \vert 0 \rangle  =
e^{iHt } 
\int 
A_N (\mathbf{p}_1, \cdots ,\mathbf{p}_N) 
e^{-i (\sum \omega_{\lambda}(\mathbf{p}_i))t} 
f_1(\mathbf{p}_1) \cdots  f_N(\mathbf{p}_N)  
d\mathbf{p}_1 \cdots d\mathbf{p}_N =
\label{e.9}
\eeq 
\beq
e^{iHt } 
\int 
A_N (\mathbf{p}_1, \cdots ,\mathbf{p}_N) 
e^{-i H_0 t} 
f_1(\mathbf{p}_1) \cdots  f_N(\mathbf{p}_N)  
d\mathbf{p}_1 \cdots d\mathbf{p}_N 
\label{e.10}
\eeq
where 
\beq
A_N (\mathbf{p}_1, \cdots ,\mathbf{p}_N) := 
\prod \left ( 
[H, \phi_{\rho} (0,\mathbf{p}_i)] - \omega_{\lambda}(\mathbf{p}_i) 
\phi_{\rho} (0,\mathbf{p}_i ) \right ) \vert 0 \rangle .
\label{e.11}
\eeq
Equation (\ref{e.9}) has the form
\beq
e^{iH t} A_N e^{-i H_0 t} \vert \mathbf{f} \rangle 
\label{e.12}
\eeq
where  $H_0 =
\sum \omega_{\lambda}(\mathbf{p}_i)$ is the
Hamiltonian for $N$ non-interacting particles of mass $\lambda$.

In this notation the Haag-Ruelle theorem, (\ref{e.3}) has the 
two-Hilbert space form: 
\beq
\lim_{t \to \pm \infty}
\Vert \vert \Psi_{\pm}(\mathbf{f})\rangle -  e^{iH t}A_{N} 
e^{-i H_0 t} \vert \mathbf{f} \rangle \Vert =0 .
\label{e.13}
\eeq

Following what is done in standard quantum mechanical 
multichannel scattering theory, wave operators are defined by
\beq
\Omega_{N\pm} \vert \mathbf{f} \rangle :=
\lim_{t \to \pm \infty}
e^{iH t}A_{N} e^{-i H_0 t} \vert \mathbf{f} \rangle 
= \vert \Psi_{\pm}(\mathbf{f}) \rangle .
\label{e.14}
\eeq
In the field theory case \cite{Ruelle:1962} these wave 
operators satisfy relativistic intertwining relations 
\beq
U(\Lambda ,a) \Omega_{N\pm} = \Omega_{N\pm} \left (\otimes U_i 
(\Lambda ,a) \right ) 
\label{e.15}
\eeq
that relate the dynamical representation of the Poincar\'e group with 
the tensor product of $n$ single-particle irreducible representations
on the $n$-particle sector of the asymptotic Fock space.

The scattering states can be expressed using the representation of the
physical Hilbert space in terms of wave functionals in section 2.  
The relevant observation is that 
\beq
\tilde{\phi}_{\rho} (p) \vert 0 \rangle 
\label{e.16}
\eeq
is a single-particle state of linear momentum $\mathbf{p}$ and
mass $\lambda$. The wave functionals (\ref{b.7}) of section 2 are vectors in
the physical Hilbert space, even though they are
expressed in terms of Euclidean test functions and the Euclidean
generating functional.  
The wave functional,
\beq
\Psi_{\lambda, j', \mathbf{p}',\mu' } [\phi] ,
\label{e.17}
\eeq
defined in the previous section,
creates a single-particle state of linear momentum $\mathbf{p}$
and mass $\lambda$.  
%

Thus, if we make the replacements  
\beq
\tilde{\phi}_{\rho} (p) \to 
\Psi_{\lambda, j', \mathbf{p}',\mu' } [\phi],
\label{e.18}
\eeq
in the two-Hilbert space Haag-Ruelle injection operator, $A_N$ (\ref{e.11}),
then it becomes the wave functional 
\beq
A_N (\mathbf{p}_1,\mu_1, \cdots \mathbf{p}_N,\mu_N )[\phi] := 
\prod \left ( 
[H, \Psi_{\lambda_i, j_i', \mathbf{p}_i',\mu_i' }] 
- \omega_{\lambda_i}(\mathbf{p}_i)
\Psi_{\lambda_i, j_i', \mathbf{p}_i',\mu_i' } 
\right ) [\phi] 
\label{e.19}
\eeq
where this functional allows for the possibility of scattering of composite
particles with higher spin.

This leads to the following formal expression for $S$- matrix elements between 
normalizable states: 
\[
S_{fi} = \langle \Psi_+ \vert \Psi_- \rangle = 
\]
\[
\lim_{t \to \infty}
\int f_1^*(\mathbf{p}_1,\mu_1)  \cdots f_M^*(\mathbf{p}_M,\mu_M)
e^{i \sum \omega_{\lambda_i}(\mathbf{p}_i)  t} \times
\]
\[ 
\langle A_M^{\dagger}(\mathbf{p}_1,\mu_1, \cdots ,
\mathbf{p}_M,\mu_M ) \vert e^{-2iHt} \vert   
A_N (\mathbf{p}'_1,\mu'_1, \cdots ,
\mathbf{p}_N',\mu_N' ) \rangle \times
\]
\beq
e^{i \sum \omega_{\lambda_i}(\mathbf{p}'_i)  t}  
f_1'(\mathbf{p}_1',\mu_1')  \cdots f_N'(\mathbf{p}_N',\mu_N')
\prod_{ij} d\mathbf{p}_i d\mathbf{p}_j'
\label{e.20}
\eeq
where the scalar product is expressed in terms of the Euclidean
generating functional.  This expression has the form
\beq
S_{fi}= \lim_{t \to \infty}\langle B(t) \vert  e^{-2iHt} \vert C(t) \rangle .
\label{e.21}
\eeq
While (\ref{e.21}) involves operators that are defined in the 
wave functional representation,  the real time evolution 
operator, $e^{-2iHt}$, is difficult to calculate in this representation.

Fortunately this quantity can be replaced by a more easily computable 
quantity using the Kato-Birman invariance principle \cite{kato:1966}
\cite{Chandler:1976}\cite{simon}\cite{baumgartl:1983}
which identifies the limits 
\beq
\Omega_{\pm} \vert \mathbf{f} \rangle :=
\lim_{t \to \pm \infty}
e^{iH t}A_{N} e^{-i H_0 t} \vert \mathbf{f} \rangle =
\lim_{t \to \pm \infty}
e^{ig(H) t}A_{N} e^{-i g(H_0) t} \vert \mathbf{f} \rangle 
\label{e.22}
\eeq
for $g(x)$ in a suitable class of admissible functions provided both
limits exist.  The content of this result is that in the large-time
limit the surviving terms correspond to situations where both
exponents oscillate in phase, which requires that both the dynamical
and asymptotic energies are the same.  Replacing $H$ and $H_0$ by
functions of $H$ resp. $H_0$ does not change this result provided the
function is increasing, with suitable smoothness.  A useful choice for
$g(x)$ that is in the class of admissible functions is
\beq
g(x) = - e^{-\beta x} \qquad \beta >0 .  
\label{e.23}
\eeq
For this choice the expressions for the wave operator becomes
\beq
\Omega_{\pm} \vert \mathbf{f} \rangle :=
\lim_{n \to \pm \infty}
e^{-in e^{-\beta H}}A_{N} e^{i n e^{-\beta  H_0}} \vert \mathbf{f} \rangle 
\label{e.23b}
\eeq
where the time parameter $t$ has been replaced by a dimensionless integer $n$.

This means that the expression (\ref{e.21}) for the $S$ matrix elements 
can be replaced by 
\[
S_{fi} = \langle \Psi_+ \vert \Psi_- \rangle = 
\lim_{n \to \infty}
\int f_1^*(\mathbf{p}_1,\mu_1)  \cdots f_k^*(\mathbf{p}_M,\mu_M)
e^{- i n e^{-\beta (\sum \omega_{\lambda_i}(\mathbf{p}_i)}}  \times
\]
\[
\times \langle A_M^{\dagger}(\mathbf{p}_1,\mu_1, \cdots ,
\mathbf{p}_M,\mu_M ) \vert e^{2in e^{-\beta H}} \vert  
A_N (\mathbf{p}'_1,\mu'_1, \cdots ,
\mathbf{p}_N',\mu_N' ) \rangle
\]
\beq
\times e^{- i n e^{i \sum \omega_{\lambda_i}(\mathbf{p}'_i)}}  
f'_1(\mathbf{p}_1',\mu_1')  \cdots f'_M(\mathbf{p}_M',\mu_M')
\prod_{ij} d\mathbf{p}_i d\mathbf{p}_j' .
\label{e.24}
\eeq
The virtue of this expression is that for large {\it fixed} $n$, 
$e^{ 2inx}$ can be 
{\it uniformly} approximated by a polynomial for $x \in [0,1]$
\beq
\vert e^{2in x} - P_{n,\epsilon} (x) \vert < 
\epsilon \quad \forall \, x \in [0,1] .
\label{e.25}
\eeq
Because the spectrum of $e^{-\beta H}$ is in the interval $[0,1]$
and the approximation is uniform
the operator satisfies the same inequality
\beq
|\Vert  e^{2in e^{-\beta H}} - P_{n,\epsilon}  (e^{-\beta H}) |\Vert < \epsilon 
\label{e.26}
\eeq
where the norm on the left is the uniform or operator norm and 
$P_{n,\epsilon}(x)$ and $\epsilon$ are the polynomial and error that 
appear in equation (\ref{e.25}).  

Compared to (\ref{e.21}) equation (\ref{e.24}) has the form 
\beq
S_{fi}= \lim_{n \to \infty}
\sum_m d_m(n) \langle B(n) \vert  e^{-\beta mH} \vert C(n) \rangle .
\label{e.27}
\eeq
where $d_m(n)$ are the coefficients of the polynomial in (\ref{e.26}). 
This is useful because matrix elements of powers of $e^{-m\beta H}$
between wave functionals $B(n)[\phi]$ and $C(n)[\phi]$
can be expressed directly in terms of the generating functional 
using (\ref{c.17}).

While time-dependent methods are not traditionally used in scattering
calculations, they have been used successfully in non-relativistic
few-body calculations \cite{kroger}.  The advantage of the above
formalism is that the entire calculation can be performed using only
Euclidean methods.

In order to calculate sharp-momentum transition matrix elements it is
necessary to use narrow wave packets.  If the transition matrix is
sufficiently smooth as a function of momentum, it will factor out of the
$S$-matrix element, allowing one to define scattering observables that
do not depend on the details of the wave packet.  This is an
assumption in the standard formulation relating time-dependent and
time-independent scattering \cite{Brenig:1959}.  For sharp initial and final wave packets
the on-shell transition matrix elements can be approximated by:
\[
\langle \mathbf{p}_1, \mu_1, \cdots \mathbf{p}_N, \mu_N 
\vert T \vert \mathbf{p}_1, \mu_1, \cdots \mathbf{p}_2, \mu_2
\rangle
\]
\beq
\approx
{\langle B \vert S \vert C \rangle -\langle B \vert C \rangle
\over -2\pi i \langle B \vert \delta ^4 (p_f-p_i) \vert C \rangle } .
\label{e.33}
\eeq 

After the wave packets are fixed the limit
$n \to \infty$ in (\ref{e.24}) can be investigated.  For a large enough $n$ 
the term in the limit, (\ref{e.24}), which has the form 
\beq
\langle B(n) \vert  e^{-2in e^{-\beta H}} \vert C(n) \rangle , 
\label{e.34}
\eeq
will be a good approximation to 
\beq
\langle B \vert S  \vert C \rangle  =
\langle B(0) \vert S  \vert C(0) \rangle .
\label{e.35}
\eeq

For this value of $n$,  $e^{-2in e^{-\beta H}}$ can then 
be uniformly approximated by a polynomial in $e^{-\beta H}$
which can be evaluated using Euclidean methods.
\beq
\langle B(n) \vert  e^{-2in e^{-\beta H}} \vert C(n) \rangle  
\approx \sum_m d_m(n)  
\langle B(n) \vert  e^{-m \beta H} \vert C(n) \rangle .
\label{e.36}
\eeq
Combining these three approximations gives an approximation to 
sharp-momentum transition matrix elements using matrix elements of
$e^{-n\beta H}$ in normalizable states as input.

Once the scattering states are known their Poincar\'e transformation
properties are determined by equation (\ref{e.15}) and the
transformation properties (\ref{d.9}) of the single-particle states.

It is useful summarize the steps needed to calculate transition-matrix
elements.  

\begin{itemize}
\item [1.] Solve the one-body problem.  These are eigenstates of the
mass-square operator with discrete eigenvalues:  $\Phi [\phi]$.  
\item [2.] Use translational and rotational covariance to 
construct $\Phi ((\lambda ,j), \mathbf{p}, \mu)  [\phi]$.
\item [3.] Choose a sufficiently narrow set of single 
asymptotic particle wave packets $f(\mathbf{p}_i ,\mu_i)$.
The width must be sufficiently narrow to factor the
transition matrix elements out of the $S$-matrix elements.

\item [4.] Construct the two-Hilbert space mappings
\beq
A_n (\mathbf{f}) := 
\prod_i\int \left (  [H,\Phi ((\lambda ,j) \mathbf{p}_i, \mu_i)]-\omega_\lambda 
(\mathbf{p}_i)\Phi ((\lambda ,j) \mathbf{p}_i, \mu_i)\right )
f(\mathbf{p}_i ,\mu_i) d\mathbf{p}_i  
e^{in e^{-\beta ( \sum \omega_\lambda 
(\mathbf{p}_i)}}    
\label{e.36b}
\eeq
\item [5.] Pick a large enough $n$.
\item [6.] Make a polynomial approximation to $e^{2inx}$ for $x\in [0,1]$
\beq
e^{2inx} \approx P_{2n,\epsilon}(x) .
\label{e.37}
\eeq
\item [7.] Calculate 
\beq
S_{fi} = \langle A (\mathbf{f}',n) 
\vert P_{2n,\epsilon}(e^{- \beta H}) \vert A (\mathbf{f}',n)\rangle
\label{e.38}
\eeq
\item [8.] Approximate $\langle \mathbf{p}_1, \mu_1, \cdots \mathbf{p}_n, \mu_n 
\vert T \vert \mathbf{p}_1, \mu_1, \cdots \mathbf{p}_2, \mu_2
\rangle$ using (\ref{e.33}).
\end{itemize}
The result can be expressed directly in terms of the
Euclidean generating functional using (\ref{c.17}). 

The discussion above assumes that the starting point is a Euclidean
generating functional.  For models the requirement that all of these
approximations converge are model assumptions that restrict properties
of the model generating functional or Green functions.  These are
reasonable requirements, since the model Green functions are modeled
after the field theoretic Green functions, which are expected to have
these properties, at least approximately.

The usual difficulties of realizing the Poincar\'e symmetry are
replaced by the requirement of finding reflection positive Euclidean
invariant Green functions or generating functionals.  It is interesting 
that the scattering matrix constructed above will be unitary, 
even if the Green function is given perturbatively.  

The proof of the existence of the wave operators in the Haag-Ruelle
formulation of scattering depends on properties of the field theory
that may be violated on approximation.  For the case of models the
existence of channel wave operators can be established by a
generalization of Cook's method\cite{cook}, which gives the following
sufficient condition for the existence of N-particle wave operators:
\beq
\int_0^{\pm \infty} \Vert ([H,A_N]-A_N H_0) e^{-i H_0t} \vert \mathbf{f} 
\rangle \Vert dt 
< \infty .
\label{e.39}
\eeq 
For the simple kinds of two-body models discussed in section VII it
is easy to see that the generalized Cook condition (\ref{e.39}) 
translates into regularity properties of the connected four-point 
function.  This is because 
\beq
\Vert ([H,A_2]-A_2 H_0) e^{-i H_0t} \vert \mathbf{f} 
\rangle \Vert^2
\label{e.40}
\eeq
is linear in the four-point Euclidean Green function and vanishes 
when the Green function is replaced by the free four-point Euclidean 
Green function.  What controls the integral in (\ref{e.40}) is the 
difference between the full and free four-point Euclidean 
Green functions, which is the connected Euclidean four-point function. 

Finally we note that even though the calculation of the scattering
observables are based on Euclidean quantities, the mechanism that
leads to the convergence of the wave operators is in-phase
oscillations, not an exponential fall-off.  The scattering states
involve strong limits and replace interpolating fields by fields that
create single-particle states out of the vacuum.  As a result, these
calculations are not subject to some of the difficulties encountered
in scattering calculations based on a Euclidean lattice
discretization\cite{Maiani}, however it is necessary to be able to
accurately compute matrix elements of $e^{-\beta H}$.

\section{Green function representation} 

While the Euclidean generating functionals provide a concise and
elegant description of the theory as well as a consistent treatment of
the few and many-body problems, the direct Green function approach of
Osterwalder and Schrader may be more appropriate for constructing
phenomenological few-body models.

In the Green function approach the representation of the physical 
Hilbert space in terms of Euclidean wave functionals is replaced 
by sequences of positive-time support functions of the form:
\beq
B[\phi] \to \langle \mathbf{\x} \vert \mathbf{f} \rangle := 
\left (
\begin{array}{c}
f_0 \\
f_1(\x_{11})\\
f_2(\x_{21},\x_{22})\\
\vdots
\end{array}
\right )
\label{f.1}
\eeq
with an inner product that is expressed in terms of multipoint
Euclidean Green functions.
\beq
\langle \mathbf{g} \vert \mathbf{f} \rangle :=
\sum_{mn}
\int d\x_1 \cdots d\x_m d\y_n \cdots d\y_1 
g_m^* (\Theta \x_1 \cdots \Theta \x_m)
G_{m+n}(\x_1,\cdots , \x_m, \y_n\, \cdots ,  \y_1) 
f_n (\y_1, \cdots, \y_n).
\label{f.2}
\eeq
In the Green function representation the support of $f_n (\y_1,
\cdots, \y_n)$ is for $0 < \y_1^0 < \y_2^0 <\cdots$.  Note that the
order of the support of the Euclidean times is identical to the order
of the fields in the corresponding Minkowski Wightman function.
Reflection positivity is the condition
\beq
\langle 
\mathbf{f} \vert \mathbf{f} 
\rangle \geq 0 
\label{f.2x}
\eeq 
and  $\langle \mathbf{g} \vert \mathbf{f} \rangle$ is the physical
quantum mechanical scalar product.

The relation between the Euclidean and Minkowski scalar products is
illustrated for the case of a free Euclidean two point function,
one for spin zero and one for spin $1/2$: 
\[
\langle 
\mathbf{f} \vert \mathbf{f} 
\rangle = 
\int f(x) G_{2}(\Theta x,y) f(y) d^4x d^4y =
\]
\[ 
= {1 \over (2 \pi)^4} \int d^4\x d^4\y d^4\p f (\x) { e^{i \p \cdot (\theta 
\x-\y)}
\over \p^2 + m^2}  f(\y) 
\]
\[
={1 \over (2 \pi)^4} \int d^4\x d^4\y d^4\p f (\x) { e^{-i \p_0 \cdot (\x_0+\y_0)
+ i \vec{\p} \cdot ( \vec{\x}-\vec{\y})}
\over  (\p^0 + i \omega_m (\vec{\p}\,))(\p^0 - i \omega_m (\vec{\p}\,))}
f(\y) 
\]
\beq
=\int d^3 \p 
{\vert g(\vec{\p}) \vert^2 \over 2\omega_m (\vec{\p} \,) }
\geq 0
\label{f.2a}
\eeq
\vfill
where 
\vfill
\beq
g(\vec{\p} ) := 
{1 \over (2\pi)^{3/2} } \int d^4y f(\y)
e^{-\omega_m (\vec{\p})y_0 -i \vec{\p} \cdot \vec{\y}}. 
\label{f.2b}
\eeq
For spin $1/2$:
\[
\langle 
\mathbf{f} \vert \mathbf{f} 
\rangle = 
\int f(x) \gamma^0 G_{2}(\Theta x,y) f(y) d^4x d^4y =
\]
\[ 
= {1 \over (2 \pi)^4} \int d^4\x d^4\y d^4\p f (\x) \,{ e^{i \p \cdot (\theta 
\x-\y)}}
\gamma^0 {m - p \cdot \gamma_e \over \p^2 + m^2}
f(\y)  
\]
\beq
=\int g^{\dagger}(\vec{\p}) {\Lambda_+ (p) \over (2 \pi)^3} 
g(\vec{\p}) d^3 \p 
\label{f.2c}
\eeq
where
\beq
\Lambda_+ (p) := {\omega_m (\vec{p}) +  
\gamma^0 \vec{\gamma} \cdot \vec{p} - m \gamma^0 
\over 2 \omega_m (\vec{p})} .
\label{f.2d}
\eeq
We see clearly that in both case the Euclidean integrals with the 
Euclidean time reversal are identical to the corresponding Minkowski 
scalar products.  The equality shows that analytic continuation is 
not required to compute the physical scalar product in the 
Euclidean representation.

In the Green-function representation the formulas for the Poincar\'e
generators are replaced by
\beq
\langle \x \vert H \vert \mathbf{f} \rangle :=
\{0 ,{\partial \over \partial
\x^0_{11}} f_1 (\x_{11}), 
\left ( {\partial \over \partial
\x^0_{21}} + {\partial \over \partial
\x^0_{22}} \right )  f_2 (\x_{21},\x_{22}), \cdots  \}
\label{f.3}
\eeq
\beq
\langle \x \vert \mathbf{P}  \vert \mathbf{f} \rangle :=
\{0 ,  - i {\partial \over
\partial \vec{\x_{11}}} f_1 (\x_{11}), 
 - i \left ( {\partial \over
\partial \vec{\x_{21}}}  + {\partial \over
\partial \vec{\x_{22}}} \right ) f_2 (\x_{21},\x_{22}), \cdots  \}
\label{f.4}
\eeq
\beq
\langle \x \vert \mathbf{J}  \vert \mathbf{f} \rangle :=
\{0 , - i \vec{\x}_{11} \times
{\partial \over \partial \vec{\x}_{11}} f_1 (\x_{11}), 
  - i\left (  \vec{\x}_{21} \times
{\partial \over \partial \vec{\x}_{21}}  + \vec{\x}_{22} \times
{\partial \over \partial \vec{\x}_{22}} 
\right  ) f_2 (\x_{21},\x_{22}), \cdots  \}
\label{f.5}
\eeq
\[
\langle \x \vert \mathbf{K}  \vert \mathbf{f} \rangle :=
\{0 , \left ( \vec{\x}_{11} 
{\partial \over \partial \x_{11}^0}- \x_{11}^0  {\partial \over
\partial \vec{\x}_{11}} \right )  f_1 (\x_{11}),
\]
\beq
\left ( \vec{\x}_{21} 
{\partial \over \partial \x_{21}^0}- \x_{21}^0  {\partial \over
\partial \vec{\x}_{21}}+ \vec{\x}_{22} 
{\partial \over \partial \x_{22}^0}- \x_{22}^0  {\partial \over
\partial \vec{\x}_{22}} \right )  f_2 (\x_{21},\x_{22}), \cdots  \}.
\label{f.6}
\eeq
For particles with spin these expression are modified as follows 
\beq
\mathbf{J} : \qquad 
\left ( - i \vec{\x}_{11} \times
{\partial \over \partial \vec{\x}_{11}} \right ) \to 
\left ( - i \vec{\x}_{11} \times
{\partial \over \partial \vec{\x}_{11}} + \vec{\Sigma} \right )
\label{f.7}
\eeq
\beq
\mathbf{K} : \qquad 
\left( \vec{\x}_{11} 
{\partial \over \partial \x_{11}^0}- \x_{11}^0  {\partial \over
\partial \vec{\x}_{11}} \right ) \to 
\left( \vec{\x}_{11} 
{\partial \over \partial \x_{11}^0}- \x_{11}^0  {\partial \over
\partial \vec{\x}_{11}}   +\vec{{\cal B}} \right ) 
\label{f.8}
\eeq
where 
\beq
\vec{\Sigma} = i \vec{\nabla}_{\phi} S(e^{{-i\over 2}\vec{\sigma}\cdot 
\vec{\phi}},  e^{{i\over 2}\vec{\sigma}^t\cdot 
\vec{\phi}})_{aa'}
\label{f.9}
\eeq
and 
\beq
\vec{{\cal B}} =  \vec{\nabla}_{\rho}S(e^{{-i\over 2}\vec{\sigma}\cdot 
\vec{\rho}},  e^{{-i\over 2}\vec{\sigma}^t\cdot 
\vec{\rho}})_{aa'} 
\label{f.10}
\eeq
Here $S(R_1,R_2)$ is a finite dimensional representations of
$SU(2)\times SU(2)$ associated with the type of field.  It is 
constructed by expressing the finite dimensional representation of the 
Lorentz group $S(\Lambda)$ in terms of $SL(2,C)$ matrices
$S(A,A^*)$ and subsequently replacing $A$ and $A^*$ by independent 
unitary matrices, $A$ and $B$.   

For the case of fermions the Euclidean time reversal operator also 
includes a factor $\gamma^0$ for each final particle.  

The formulas summarized in this section are discussed in more
detail in \cite{Wessels:2003af}.
 
\section{Few-body models}

A typical application where relativistic few-body methods are used is
elastic nucleon-nucleon scattering.  This is normally treated using
the inhomogeneous Bethe-Salpeter equation.  We outline the formulation
of this problem in the Euclidean quantum mechanical representation.

We consider a model Green function of the form
\beq
\Theta S \to 
\left (
\begin{array}{cccc}
0 & 0 & 0 
& \cdots \cdots\\
0 & S_2 (\Theta \x_{11},\x_{12}) 
& 0 & 
\\
0 & 
0 & S_4
(\Theta \x_{21},\Theta \x_{22},\x_{23},\x_{24}) &
0  \\
\vdots & 0 & 0  &    
\ddots
\end{array}
\right ).
\label{g.1}
\eeq
For this model we assume that only $S_2$ and $S_4$ are non-zero.
Furthermore we assume that $S_2$ and $S_4$ are related by cluster 
properties:
\beq
S_4 = S_2 S_2 + S_c = S_0 + S_c .   
\label{g.2}
\eeq
The Euclidean Bethe Salpeter kernel is defined by  
\beq
S_4^{-1} - S_0^{-1} = - K .
\label{g.3}
\eeq
The structure of $S_2$ is determined by covariance up to an unknown
Lehmann weight.  If the weight is a delta function in the mass then
this is a free field Euclidean Green function.  In this case the
one-body solutions that are needed to formulate the scattering problem
are trivial.  If the Lehmann weight also includes some continuous
spectrum then it is necessary to solve a one-body problem to formulate
the scattering asymptotic condition.   To do this we take an orthonormal 
set of positive-time test functions and use the Gram-Schmidt method to 
construct an orthonormal set
\beq
\langle f_n \vert f_m \rangle =
\int d\x d\y f_n^* (\Theta \x )\gamma^0 S_2 (\x-\y) f_m (\y) = \delta_{mn}.
\label{g.4}
\eeq
Because the invariant Minkowski Green function is 
defined with a Dirac conjugate field rather than a 
Hilbert space adjoint, the $\gamma^0$ needs to be eliminated
from $S_2$ to get the continuation to the Wightman function 
that serves as the kernel of the Hilbert space scalar product.  
This is achieved by including $\gamma^0$ as the spinor 
part of the $\Theta$ operator:

In this basis one-body solutions have the form
\beq
\vert \lambda \rangle = \sum_n c_n \vert \mathbf{f}_n \rangle 
\label{g.5}
\eeq
where $c_n$ and $\lambda$ are determined solving the eigenvalue problem
for discrete $\lambda^2$:
\beq
\lambda^2 c_n = 
\sum_m ({\partial^2 \over \partial \tau^2} 
- {\partial^2 \over \partial \mathbf{a}^2})
(f_n \gamma^0 \Theta S_2 
f_{m,I,(\tau,\mathbf{a})})_{\tau=\mathbf{a}=0} 
c_m  .
\label{g.6}
\eeq
In this case the Euclidean two-point Green function has the form 
\beq
S_{2}(\x-\y) := 
{1 \over (2 \pi)^4} \int {d^4 \p} \rho(m) dm
{m - p \cdot \gamma_e \over \p^2 + m^{ 2}} e^{i \p \cdot (\x-\y)}
\label{g.7}
\eeq
where 
\beq
i \gamma_{0e} = \beta = \gamma^0 = -\gamma_0; \qquad \gamma^i_e = \gamma^i .
\label{g.8}
\eeq 
and 
\beq
\rho (m) = \rho_m\delta (m-\lambda) + \rho_c (m).
\label{g.9}
\eeq
The matrix elements have the form
\[
(f,\Theta \gamma^0 S_{2}f) 
\]
\[ 
= {1 \over (2 \pi)^4} \int d^4\x d^4\y d^4\p dm f (\x) \,{ e^{i \p \cdot (\Theta 
\x-\y)}}
\gamma^0 {m - p \cdot \gamma_e \over \p^2 + m^2} \rho(m)
f(\y)  
\]
\beq
=\int g^{\dagger}(\mathbf{p},m) {\Lambda_+ (\mathbf{p},m) \over (2 \pi)^3} \rho(m)
g(\vec{\p},m) d\mathbf{p} dm 
\label{g.10}
\eeq
where
\beq
\Lambda_+ (p) := {\omega_m (\mathbf{p}) +  
\gamma^0 \vec{\gamma} \cdot \mathbf{p} - m \gamma^0 
\over 2 \omega_m (\mathbf{p})} 
\label{g.11}
\eeq
is the positive energy Dirac projector and
\beq
g(\mathbf{p},m) := \int d^4\x \, 
e^{- \omega_m (\mathbf{p}) \x_0 
- i \mathbf{\p} \cdot \mathbf{\x}} f(\x) . 
\label{g.12}
\eeq

Single particle eigenstates of mass, linear momentum and spin are 
constructed from the mass eigenstates (\ref{g.6})
\beq
\psi_\lambda (\x) = \sum_n c_n f_n (\x )
\eeq
using 
\beq
\psi_\lambda (\mathbf{p},\mu) =
\sum_n c_n {1 \over (2 \pi)^{3/2} } \int \sum_{\nu}   
\mathbf{f}_n(\tau,R^{-1} \mathbf{x}- \mathbf{a}) 
e^{-i R^{-1}\mathbf{p}\cdot \mathbf{x}} d\mathbf{a}
D^{j*}_{\mu \nu} (R) dR 
\label{g.21}
\eeq

The Haag-Ruelle operators $A(\mathbf{p},\mu)$ are  
\beq
A(\mathbf{p},\mu)[\x] = 
{1 \over (2\pi)^{3/2}}
\int d\mathbf{p} 
[-{\partial \over \partial \beta} -\omega_\lambda (\mathbf{p})]
\psi_\lambda (\mathbf{p},\mu)[\tau-\beta,\mathbf{x}-\vec{a}]_{\vert_{\beta=\mathbf{a}=0}}
\label{g.23}
\eeq
The scattering asymptotic states of interest are two-body states.
$S$-matrix elements in normalizable states can be computed using the
methods discussed in section 5.  Equation (\ref{e.33}) is replaced by:
\[
\langle g \vert S  \vert f \rangle = 
\]
\[
\lim_{n \to  \infty} \sum_{d_n} \int g(\mathbf{p}_1',\mu_1')
g(\mathbf{p}_2',\mu_2') e^{ine^{-\beta (\omega_\lambda
    (\mathbf{p}_1')+ \omega_\lambda (\mathbf{p}_2'))}}
A^{\dagger}(\mathbf{p}'_1,\mu_1')[\Theta \x_2']
A^{\dagger}(\mathbf{p}'_2,\mu'_2)[\Theta \x_1'] \gamma_1^0 \gamma^0_2
\]
\[
\times
S_4 (\x_2',\x_2';\x_1, \x_2) A(\mathbf{p}_1,\mu_1)[\x_1-2n\beta]
A(\mathbf{p}_2,\mu_2)[\x_2-2n\beta] e^{ine^{-\beta (\omega_\lambda
    (\mathbf{p}_1')+ \omega_\lambda (\mathbf{p}_2'))}} \times
\]
\beq
g(\mathbf{p}_2,\mu_2) g(\mathbf{p}_1,\mu_1) d \mathbf{p}_1
d\mathbf{p}_2 d \mathbf{p}_1'd \mathbf{p}_2' d\x_2 d\x_2 d\x_1' d\x_2'
\label{g.24}
\eeq
Here the reflection positivity is limited to requiring 
that $\gamma^0_1 \gamma^0_2\Theta S_4$ is non-negative 
on products of positive-time test functions. 

This illustrates how the approximations discussed above can be
implemented in a few-body setting.  We note that even if
$S_4$ is calculated perturbatively, the resulting approximate 
$S$ matrix will be unitary.

\section{Scattering test}

The scattering computations outlined above and in section V are based
on the convergence of a sequence of three approximations that are
performed in a prescribed order.  While they should in principle
converge for suitable model Green functions, that does not imply that
the approximations can be sufficiently well-controlled to give
converged predictions for reactions at the few GeV energy scale of
interest.  Since to the best of our knowledge this approach to
scattering, i.e. computing sharp-momentum transition matrix elements
using matrix elements of $e^{-n\beta H}$ in normalizable states as
input, has not even been tested in non-relativistic models, we discuss
the implementation of this method in an exactly solvable
non-relativistic model.  This has the advantage that all of the
approximations can be compared to exact results, and the accuracy of
the proposed method can be determined precisely.  We consider a
non-relativistic Hamiltonian with a separable potential that has the
range and strength of a nucleon-nucleon interaction.  The range is
fixed by the pion mass while the strength is adjusted to bind two
nucleons with the deuteron binding energy.

The the interaction is taken as a Yamaguchi interaction,
with Hamiltonian
\beq
H= \mathbf{k}^2 /m - \vert g \rangle \lambda \langle g \vert 
\qquad
\langle \mathbf{k} \vert g \rangle = g(\mathbf{k}) = {1 \over m_{\pi}^2 + \mathbf{k}^2} .
\eeq
The transition matrix elements are 
\beq
\langle \mathbf{k} \vert t \vert \mathbf{k}' \rangle =
g(\mathbf{k})
\tau ({k^{\prime 2} \over  m_n} + i0^+)
g(\mathbf{k}')
\eeq
with 
\beq
\tau ({k^{\prime 2} \over  m_n} + i0^+) = 
- {\lambda \over 1 + { m_n \lambda \pi^2 \over m_\pi}{1 \over 
(im_\pi + k')^2}} 
\eeq
where $m_n=.94$GeV, $m_\pi=.14$GeV, $e_b=-2.24$MeV and the coupling
constant is determined from these parameters by 
\beq
\lambda={m_\pi \over m_n \pi^2} (m_\pi+\sqrt{-m_n e_b})^2 .
\eeq
To test the approximations we calculate sharp-momentum transition
matrix elements using matrix elements of $e^{-\beta H}$ evaluated
between normalizable states using the methods outlined in section
five.  While the eigenstates of this model can be computed exactly, we
had to use the spectral expansion of the Hamiltonian to compute matrix
elements of $e^{-\beta H}$.  While this is a complicated construction
in the non-relativistic case, it is replaced by an elementary
quadrature (\ref{c.17}) in the Euclidean case.

The first approximation is to extract sharp-momentum transition matrix
elements using sufficiently narrow wave packets in equation
(\ref{e.33}).  For this model exact expressions are available for both
the $S$ operator and transition matrix.  For our test problem we
choose Gaussian wave packets in the relative momenta of the form
\beq
\phi (k)= N e^{-\alpha (k-k_0)^2}= N e^{-(k-k_0)^2/k_w^2}   
\eeq
where $N$ is a normalization constant and $k_0$ is the on-shell
momentum.  We do not choose a particular direction because the
interaction is pure $s$-wave, and the on-shell transition matrix 
elements at a given energy are given by a single complex number.

Sharp on-shell transition matrix elements computed exactly and
approximately from $S$-matrix elements using (\ref{e.33}) are compared
as a function of the width, $k_w$ of the wave packets.  The wave
packet widths were determined by the requirement that the approximate
transition matrix elements agree with the exact transition matrix
elements to an accuracy of less that 0.1\%.  The results are shown in
Table 1 as a function of the relative momentum.  The first column of
Table 1 shows the on-shell momentum $k_0$ (center of the Gaussian).
The second column shows the value of $\alpha$ used to get the error
shown in the fourth column.  All of the errors in column four round up
to $.1\%$.  The third column lists $k_w:=1/\sqrt{\alpha}$ for each
value of $k_0$ and the last column is the dimensionless ratio
$k_w/k_0$.  The values of $\alpha$ in column 3 are used in all of the
calculations in this section.
\vbox{
\begin{center}
Table 1
\end{center}
\begin{center}
\begin{tabular}{lllll}
\hline
$k_0$ & $\alpha$ &  $k_w$  &    \% error &  $k_w/k_0$ \\
\hline
[GeV] & [GeV$^{-2}]$ & [GeV] &   &  \\
\hline
0.05 &325000	&0.00175412	&0.1     &0.035 \\
0.1  &105000	&0.00308607	&0.1     &0.030 \\
0.2  &26000	&0.00620174	&0.1     &0.031 \\
0.3  &10500	&0.009759	&0.1     &0.032 \\
0.4  &5100	&0.0140028	&0.1     &0.035 \\
0.5  &3000	&0.0182574	&0.1     &0.036 \\
0.6  &2000	&0.0223607	&0.1     &0.037 \\
0.7  &1350	&0.0272166	&0.1     &0.038 \\
0.8  &1000	&0.0316228	&0.1     &0.039 \\
0.9  &750	&0.0365148	&0.1     &0.040 \\
1.0  &600	&0.0408248	&0.1     &0.040 \\
1.1  &475	&0.0458831	&0.1     &0.041 \\
1.2  &400	&0.05	        &0.1     &0.041 \\
1.3  &330	&0.0550482	&0.1     &0.042 \\
1.4  &290	&0.058722	&0.1     &0.041 \\
1.5  &250	&0.0632456	&0.1     &0.042 \\
1.6  &210	&0.0690066	&0.1     &0.043 \\
1.7  &190	&0.0725476	&0.1     &0.042 \\
1.8  &170	&0.0766965	&0.1     &0.042 \\
1.9  &150	&0.0816497	&0.1     &0.042 \\
2.0  &135	&0.0860663	&0.1     &0.043 \\
\end{tabular}
\end{center}
}
The last column of Table 1 suggests that a .1\% error will generally
be obtained if this width is less than 3\% of the on-shell momentum.
This property holds over a wide range of momenta in this model.  This
is a simple transition operator so one may anticipate narrower wave
packets are needed for more realistic models.  This approximation can
be improved by decreasing the width of the wave packet; it is the
largest source of error in the calculations.  In any realistic
calculation it does not have to be better than the experimental
resolution.

The second step is to approximate $S$-matrix elements in these
Gaussian wave packets using equation (\ref{e.24}).  It
is important to first pick the wave packets because the $n$ 
value needed for convergence depends on the width of the 
wave packet.  The approximate quantities are 
\beq
\langle \phi \vert S_n \vert \phi \rangle :=
\int d\mathbf{k} d\mathbf {k}' \phi(\mathbf{k}) 
e^{- i n e^{-\beta k^2/m}}\langle \mathbf{k}e \vert
e^{-2in e^{-\beta H}} \vert \mathbf{k}' \rangle     
e^{- i n e^{-\beta k^{\prime 2}/m}} \phi(\mathbf{k}')
\eeq
In Tables 2-8 these quantities are computed using the spectral
expansion for $H$.  In these calculations the bound-state contribution
is not included because it vanishes in the large $n$ limit.
Tables 2-8 show the real and imaginary parts of matrix elements
$\langle \phi \vert( S_n-I) \vert \phi \rangle$ for different values
of $n$ for $k_0=50,100,200,500,1000,1500, 2000$ MeV.  The exact value is 
given at the
bottom of each table.  Table 9 shows the values of $\beta$, the
product $k_0\times \beta$ and the $n$-values used in our final
calculations.  Table 9 suggests that $\beta$ should be chosen so
$k_0\times \beta$ is of order unity.  Except for the $k_0=50$MeV case, 
$n=250$ or more gives errors for the $n$-limits that are smaller than 
the errors made in the factorization approximation, (\ref{e.33}).  

The $n$ dependence of the real and imaginary part of matrix elements of
$\langle \phi\vert (S-I) \vert \phi \rangle$, computed using  
(\ref{e.24}), are plotted as a function of $n$ for different values of
$k_0$ in figures 1-10.  Figures 11 and 12 show how fast the
neglected bound state contribution to the spectral expansion
fall off with $n$ for $k_0=1$ GeV. 

\vbox{
\begin{center}
Table 2: $k_0=50$[MeV], $\alpha=325000[$GeV$^{-2}]$  
\end{center}
\begin{center}
\begin{tabular}{lll}
\hline
n & Re $\langle \phi \vert (S_n-I) \vert \phi \rangle$ & Im $\langle \phi \vert (S_n-I) \vert \phi \rangle$ \\
\hline
50 &-7.62976513315350e-1 & -1.52406978178214e-1 \\
100 & -1.33113144491104e+0 & -2.75546806155677e-1 \\
150 & -1.67324498184421e+0 &  -3.50392186517949e-1 \\
200 &-1.83391449191883e+0 & -3.86136590065981e-1 \\
250 &-1.89273779093641e+0 & -3.99389077283171e-1 \\
300 &-1.90951807884485e+0 & -4.03228425149703e-1 \\
350 &-1.91324926322936e+0 & -4.04093826445785e-1 \\
400 &-1.91389545311571e+0 & -4.04245969269804e-1 \\
450 & -1.91398265491004e+0 & -4.04266741575048e-1 \\
500 &-1.91399177708580e+0 & -4.04268944214980e-1 \\
550 &-1.91399249452470e+0 & -4.04269131167755e-1 \\
600 &-1.91399253056843e+0 & -4.04269144047622e-1 \\
650 &-1.91399252976074e+0 & -4.04269145893613e-1 \\
\hline
ex  &-1.91399253060872e+0 &-4.04269147714400e-1 \\   
\hline
\end{tabular}
\end{center}
}        
           
\vbox{
\begin{center}
Table 3: $k_0=100$[MeV], $\alpha=105000[$GeV$^{-2}]$  
\end{center}
\begin{center}
\begin{tabular}{lll}
\hline
n & Re $\langle \phi \vert (S_n-I) \vert \phi \rangle$ & Im $\langle \phi \vert (S_n-I) \vert \phi \rangle$ \\
\hline
50  &-8.73395186664514e-1 &4.95616337213744e-1 \\ 
100 &-1.34576615227520e+0 &7.59199494502660e-1 \\
150 &-1.49091126760062e+0 &8.39700869213905e-1 \\
200 &-1.51566533604846e+0 &8.53352070852317e-1 \\
250 &-1.51799902547669e+0 &8.54631681615040e-1 \\
300 &-1.51811943431498e+0 &8.54697554653043e-1 \\
350 &-1.51812278309620e+0 &8.54699376334219e-1 \\
400 &-1.51812288480017e+0 &8.54699423334768e-1 \\
450 &-1.51812290596551e+0 &8.54699435479409e-1 \\
500 &-1.51812290955424e+0 &8.54699438102000e-1 \\
550 &-1.51812290968227e+0 &8.54699438639971e-1 \\
600 &-1.51812288857871e+0 &8.54699427676778e-1 \\
650 &-1.51812275938123e+0 &8.54699356334623e-1 \\
\hline   
ex  &-1.51812291315971e+0 &8.54699438329052e-1 \\
\hline
\end{tabular}
\end{center}
}        

\vbox{
\begin{center}
Table 4: $k_0=200$[MeV], $\alpha=26000$[Gev$^{-2}$]  
\end{center}
\begin{center}
\begin{tabular}{lll}
\hline
n & Re $\langle \phi \vert (S_n-I) \vert \phi \rangle$ & Im $\langle \phi \vert (S_n-I) \vert \phi \rangle$ \\
\hline
 50 &-2.08408481834932e-1 &4.56550768265380e-1 \\
100 &-3.11945696198071e-1 &6.85279276148059e-1 \\
150 &-3.38623394392403e-1 &7.44641333032490e-1 \\
200 &-3.42127100784575e-1 &7.52475859236454e-1 \\
250 &-3.42359208499266e-1 &7.52997722504423e-1 \\
300 &-3.42366821259122e-1 &7.53015087494957e-1 \\
350 &-3.42366956571344e-1 &7.53015382645461e-1 \\
400 &-3.42366963936415e-1 &7.53015400508140e-1 \\
450 &-3.42366965122021e-1 &7.53015404697748e-1 \\
500 &-3.42366965247355e-1 &7.53015405389574e-1 \\
550 &-3.42366962667700e-1 &7.53015400608940e-1 \\
600 &-3.42366938358008e-1 &7.53015348812648e-1 \\
650 &-3.42366884413759e-1 &7.53015247619997e-1 \\
\hline
ex & -3.42366967477707e-1 &7.53015410457076e-1 \\
\hline
\end{tabular}
\end{center}
}        

\vbox{
\begin{center}
Table 5: $k_0=500$[MeV], $\alpha=3000$[Gev$^{-2}$]  
\end{center}
\begin{center}
\begin{tabular}{lll}
\hline
n & Re $\langle \phi \vert (S_n-I) \vert \phi \rangle$ & Im $\langle \phi \vert (S_n-I) \vert \phi \rangle$ \\
\hline
50  & -5.93330385580271e-3 & 9.65738963854834e-2 \\ 
100 & -7.12681692349637e-3 & 1.18415504225673e-1 \\ 
150 & -7.18129565909453e-3 & 1.19491569949089e-1 \\ 
200 & -7.18179706798405e-3 & 1.19502423190014e-1 \\ 
250 & -7.18179794641838e-3 & 1.19502444469971e-1 \\ 
300 & -7.18179794670870e-3 & 1.19502444477783e-1 \\ 
350 & -7.18179794671048e-3 & 1.19502444477784e-1 \\ 
400 & -7.18179794671081e-3 & 1.19502444477784e-1 \\ 
\hline                                              
ex  & -7.18179797016073e-3 & 1.19502444795275e-1 \\
\hline
\end{tabular}
\end{center}
}
 
\vbox{
\begin{center}
Table 6: $k_0=1$[GeV], $\alpha=600$[Gev$^{-2}$]  
\end{center}                                           
\begin{center}
\begin{tabular}{lll}
\hline
n & Re $\langle \phi \vert (S_n-I) \vert \phi \rangle$ & Im $\langle \phi \vert (S_n-I) \vert \phi \rangle$ \\
\hline 
50  &-1.47024820732811e-4 &1.55922557816223e-2 \\ 
100 &-1.62726188649875e-4 &1.79713868930101e-2 \\ 
150 &-1.62967714125273e-4 &1.80219591282450e-2 \\ 
200 &-1.62968113934903e-4 &1.80220978916775e-2 \\ 
250 &-1.62968113982642e-4 &1.80220979403700e-2 \\ 
300 &-1.62968113982642e-4 &1.80220979403721e-2 \\ 
350 &-1.62968113982753e-4 &1.80220979403720e-2 \\ 
400 &-1.62968113982975e-4 &1.80220979403718e-2 \\ 
\hline                                            
ex  &-1.62968113982742e-4 &1.80220979403858e-2 \\
\hline
\end{tabular}
\end{center}
}

\vbox{
\begin{center}
Table 7: $k_0=1.5$[GeV], $\alpha=250$[GeV$^{-2}$]  
\end{center}
\begin{center}
\begin{tabular}{lll}
\hline
n & Re $\langle \phi \vert (S_n-I) \vert \phi \rangle$ & Im $\langle \phi \vert (S_n-I) \vert \phi \rangle$ \\
\hline
50   &-1.40242356887477e-5 &4.66175982621713e-3 \\ 
100  &-1.54430995201738e-5 &5.52235009412764e-3 \\ 
150  &-1.54679181726403e-5 &5.55112445776045e-3 \\ 
200  &-1.54679281958447e-5 &5.55130663695727e-3 \\ 
250  &-1.54679280390813e-5 &5.55130688968718e-3 \\ 
300  &-1.54679280390813e-5 &5.55130688978369e-3 \\ 
350  &-1.54679280394143e-5 &5.55130688978374e-3 \\ 
400  &-1.54679280388592e-5 &5.55130688978377e-3 \\ 
\hline                                             
ex   &-1.54679280242191e-5 &5.55130688386830e-3 \\
\hline
\end{tabular}
\end{center}
}

\vbox{
\begin{center}
Table 8: $k_0=2.0$[GeV], $\alpha=135$[GeV$^{-2}$]  
\end{center}
\begin{center}
\begin{tabular}{lll}
\hline
n & Re $\langle \phi \vert (S_n-I) \vert \phi \rangle$ & Im $\langle \phi \vert (S_n-I) \vert \phi \rangle$ \\
\hline
50  & -2.60094316473225e-6 & 1.94120750171791e-3 \\ 
100 & -2.82916859895010e-6 & 2.35553585404449e-3 \\
150 & -2.83171624670953e-6 & 2.37471383801820e-3 \\ 
200 & -2.83165946257657e-6 & 2.37492460997990e-3 \\ 
250 & -2.83165905312632e-6 & 2.37492527186858e-3 \\ 
300 & -2.83165905257121e-6 & 2.37492527262432e-3 \\ 
350 & -2.83165905190508e-6 & 2.37492527262493e-3 \\
400 & -2.83165905234917e-6 & 2.37492527262540e-3 \\
\hline                                             
ex  & -2.83165905227843e-6 & 2.37492527259701e-3 \\
\hline
\end{tabular}
\end{center}
}

\vbox{
\begin{center}
Table 9: Parameters  
\end{center}
\begin{center}
\begin{tabular}{llll}
\hline
k0 [GeV] & $\beta[$GeV$^{-1}$] & $k0\times\beta$ &  $n$ \\
\hline
0.05 &	80.0    &  4.0 &  630  \\
0.1  &	40.0    &  4.0 &  450  \\
0.2  &	10.0    &  2.0 &  470  \\
0.3  &	5.0     &  1.5 &  330  \\
0.4  &	4.0     &  1.6 &  235  \\
0.5  &	3.0     &  1.5 &  205  \\
0.6  &	2.5   &  1.5 &  225  \\
0.7  &	1.6   &  1.2 &  200  \\
0.8  &	1.4   &  1.12&  200  \\
0.9  &	1.05  &  .945&  190  \\
1.0  &	1.0     &  1.0 &  200  \\
1.1  &	0.95  &  1.045 & 200  \\
1.2  &	0.9   &  1.08 & 200  \\
1.3  &	0.85  &  1.105 & 200  \\
1.4  &	0.7   &  0.98 & 200  \\
1.5  &	0.63  &  0.945 & 200  \\
1.6  &	0.57  &  0.912 & 200  \\
1.7  &	0.5   &  0.85 & 200  \\
1.8  &	0.45  &  0.81 & 200  \\
1.9  &	0.42  &  0.798 & 200  \\
2.0  &	0.4   &  0.8 & 200  \\
\hline                
\end{tabular}           
\end{center}
}

The third approximation is the polynomial approximation to
$e^{i2n e^{-\beta H}}$.  In this application we shift
the potential by a constant (the binding energy) so the 
spectrum of $e^{-\beta H}$ is strictly between zero and one.
Then it is only necessary to find a polynomial approximation to
$e^{inx}$ for $x \in [0,1]$.

The polynomial approximation is made using the Chebyshev expansion:
\beq 
f(y) \approx P_N(y) := {1 \over 2} c_0 T_0 (y) + \sum_{k=1}^N c_k T_k (y) 
\label{e.28}
\eeq
\beq
c_j = {2 \over N+1} \sum_{k=1}^N f( \cos({2k-1 \over N+1}{\pi
\over 2})) \cos(j {2k-1 \over N+1}{\pi \over 2}). 
\label{e.29}
\eeq
The Chebyshev expansion
is designed for functions on the interval $[-1,1]$.  To
make full use of the information in this approximation 
we approximate $e^{-2inx} = f(2x-1)$ by the polynomial
approximation to $f(y)$. This gives the polynomial
\beq 
f(2 e^{-\beta H'}-1 ) \approx {1 \over 2} c_0 T_0 (
2e^{-\beta H'}-1) + \sum_{k=1}^N c_k T_k (2e^{-\beta H'}-1)
= P_N ( e^{-\beta H'})
\label{e.30}
\eeq
where $H'=H+e_b$. It follows that 
\beq
\vert e^{2inx} - P_N(x) \vert < 2 {{n}^{N+1}\over (N+1)!} 
\qquad
|\Vert e^{2ine^{-\beta H'}} - P_N(e^{-\beta H'}) |\Vert < 2 {{n}^{N+1}\over (N+1)!} 
\label{e.32}
\eeq
Chebyshev polynomials are know to be good approximations to
the best uniform polynomial approximation \cite{oxford2004}. 
Table 10 shows polynomial approximations to $e^{inx}$ for different 
values of $x\in [0,1]$, and $n$.  The errors are between 
$10^{-11}$ and $10^{-13}$ for the degree of the polynomial only slightly 
above $n$.
\vbox{
\begin{center}
Table 10: Polynomial convergence
\end{center}
\begin{center}
\begin{tabular}{llll}
\hline
x     &  n  &    deg   &    poly error \% \\
\hline
0.1  &	 200  &  150 &	 7.939e+00     \\
0.1  &   200  &  200 &   3.276e+00    \\
0.1  &   200  &  250 &   1.925e-11  \\
0.1  &   200  &  300 &   4.903e-13  \\
\hline
0.1  &   630  &  580 &   3.573e+00    \\
0.1  &   630  &  630 &   2.069e+00    \\
0.1  &   630  &  680 &   5.015e-08   \\
0.1  &   630  &  700 &   7.456e-11   \\
\hline                         
0.5  &   200  &  150 &   1.973e-13  \\
0.5  &   200  &  200 &   1.627e-13  \\
0.5  &   200  &  250 &   3.266e-13  \\
\hline0.5  &   630  &  580 &   1.430e-14  \\
0.5  &   630  &  630 &   3.460e-13  \\
0.5  &   630  &  680 &   9.330e-13  \\
\hline                
0.9  &   200  &  150 &   7.939e+00      \\
0.9  &   200  &  200 &   3.276e+00      \\
0.9  &   200  &  250 &   1.950e-11  \\
0.9  &   200  &  300 &   9.828e-13  \\
\hline
0.9  &   630  &  580 &   3.573e+00  \\
0.9  &   630  &  630 &   2.069e+00  \\
0.9  &   630  &	 680 &   5.015e-08  \\
0.9  &	 630  &  700 &   7.230e-11  \\
\hline
\end{tabular}
\end{center}
}
Table 10. suggests that a polynomial with degree 10-20\% larger than $n$ 
is needed for convergence to one part in $10^{12}$ ($10^{-10}$\%).

All three approximations are combined to get the approximations to the
on-shell transition operator for incident momenta between 50 MeV and 2
GeV.  These results are displayed in table 11.  The errors are all
better than .1\%.  The only significant source of error is the
approximate factorization of the sharp momentum matrix element from
the $S$ matrix elements.  This can be made as small as desired by
choosing sufficiently narrow wave packets, although there is no need
to improve accuracy to better than experimental resolution.
\vbox{
\begin{center}
Table 11: Final calculation
\end{center}
\begin{center}
\begin{tabular}{llll}
\hline
$k0$ & Real T & Im T & \% error \\
\hline
0.05  &   2.18499e-1    &  -1.03160e+0	          &     0.0982   \\
0.1   &  -2.30337e-1    &  -4.09325e-1	          &     0.0956   \\
0.2   &  -1.01512e-1    &  -4.61420e-2	          &     0.0981   \\
0.3   &  -3.46973e-2    &  -6.97209e-3	          &     0.0966   \\
0.4   &  -1.39007e-2    &  -1.44974e-3	          &     0.0997   \\
0.5   &  -6.44255e-3    &  -3.86459e-4	          &     0.0986   \\
0.6   &  -3.34091e-3    &  -1.24434e-4	          &     0.0952   \\
0.7   &  -1.88847e-3    &  -4.63489e-5           &     0.0977   \\
0.8   &  -1.14188e-3    &  -1.93605e-5           &     0.0965   \\
0.9   &  -7.28609e-4    &  -8.86653e-6           &     0.0982   \\
1.0   &  -4.85708e-4    &  -4.37769e-6           &     0.0967   \\
1.1   &  -3.35731e-4    &  -2.30067e-6           &     0.0987   \\
1.2   &  -2.39235e-4    &  -1.27439e-6           &     0.0968   \\
1.3   &  -1.74947e-4    &  -7.38285e-7           &     0.0985   \\
1.4   &  -1.30818e-4    &  -4.44560e-7           &     0.0955   \\
1.5   &  -9.97346e-5    &  -2.76849e-7           &     0.0956   \\
1.6   &  -7.73390e-5    &  -1.77573e-7           &     0.0992   \\
1.7   &  -6.08794e-5    &  -1.16909e-7           &     0.0964   \\
1.8   &  -4.85672e-5    &  -7.87802e-8           &     0.0956   \\
1.9   &  -3.92110e-5    &  -5.42037e-8           &     0.0967   \\
2.0   &  -3.20000e-5    &  -3.80004e-8           &     0.0966   \\
\hline                                                                           
\end{tabular}
\end{center}
}
These tests suggest that this method can be applied to compute 
scattering observables at the few GeV scale. 

\section{summary and conclusion} 

In this paper we introduced a formulation of relativistic quantum
mechanics that uses Euclidean Green functions or generating
functionals as dynamical input.  The motivation for this approach is
to formulate few-body models at the few GeV scale.  The models are
quantum mechanical which means that they are formulated in terms of
linear operators on a model Hilbert space.  They can be treated using
standard quantum mechanical methods.  The advantages of this
framework over some conventional treatments of relativistic quantum
mechanics is that there is a formal relation to Euclidean Lagrangian
field theory.  Specifically, the quantum theory discussed in this
paper becomes the quantum mechanical formulation of the field theory when
the model Green functions are replaced by the Green functions of
the field theory.  A second advantage of this formalism is that all
calculations can be performed in the Euclidean domain, without using
analytic continuation, however the quantities computed are ordinary
Minkowski scalar products of normalizable vectors.  This permits the
use of ordinary Minkowski-space methods.

The structure of the physical Hilbert space was discussed in section 1
in terms of a Euclidean generating functional for a scalar field
theory.  Methods to model generating functionals in terms of connected
Euclidean Green functions were also discussed.  Representations in
terms of Euclidean Green functions were given in section 6.  Examples
illustrating the equivalence of the Hilbert space inner product
expressed in terms of Euclidean Green functions and the standard
Minkowski inner product were given in equations (\ref{f.2a}) and
(\ref{f.2b}).

The Poincar\'e Lie algebra is realized by interpreting the real
Euclidean group as a complex subgroup of the complex Lorentz group on
the physical Hilbert space.  Self-adjoint generators that satisfy the
Poincar\'e commutation relations are extracted by considering
infinitesimal Euclidean transformations which become infinitesimal
complex Lorentz transformations on the physical Hilbert space.
Expressions for the generators were given both in terms of the
generating functional in section 3 and directly in terms of the Green
functions in section 6.  

In section four the generators introduced in sections three and six
are used to construct a mass operator whose point eigenstates
correspond to particles.  The translation and rotation operators
introduced in sections three and six are used to construct operators that
create single particle states of a given sharp momentum and spin out
of normalizable mass eigenstates.  These states necessarily transform
irreducibly under the Poincar\'e group if they non-degenerate.

The operators that create single-particle states are used to construct
two-Hilbert space injection operators in the Haag-Ruelle formulation
of scattering.  The existence of the wave operators can be checked
using a generalization of the standard Cook's method \cite{cook} used
in non-relativistic quantum mechanics.  The two-Hilbert space wave
operators are computed using time-dependent methods with the
Kato-Birman invariance principle to reduce the computation to the
evaluation matrix elements of polynomials in $e^{-\beta H}$.  The
feasibility of this method for computing scattering observables was
established using an exactly solvable non-relativistic test model.
This model demonstrated that it is possible to perform accurate
calculations over a wide range of energies using this method.
Heuristic guidelines for choosing the ``time'' $n$, temperature
$\beta$, and the degree of the polynomial were established in the
context of this model.  Both the ''time'' limit and polynomial
approximations were shown to be very accurate.  Even though the input
to the scattering theory involves Euclidean quantities, scattering
emerges because the scattering states asymptotically oscillate in
phase with free-particle states as the ``time-paramenter'', $n$, gets
large.  The limits do not depend on the exponential fall-off that is
used in lattice calculations.  This means that calculations can be
performed at relativistic energies without the complications that
arise in alternative formulations of scattering involving Euclidean
quantities \cite{Maiani}.  Since the output of these calculations are
wave operators, the relativistic intertwining properties provide a
mechanism for performing finite Poincar\'e transformations on all
scattering states.  In addition, this also means the even perturbative
approximations to the Green function or generating functional will
yield unitary, Poincar\'e invariant approximate $S$-matrices.

Real model calculations are done using Euclidean Green functions.  In
principle the model Green functions should be Euclidean covariant,
real, satisfy cluster properties and be reflection positive.  Taken
together these are restrictive conditions.  While good models should
satisfy these conditions, some models may fail to satisfy some of
these conditions exactly.  One may still get good approximations if
small operators are responsible for the difference between the exact
and approximate Green functions.  These questions require additional
study.  A relevant observation is that because all of the calculations
are done without analytic continuation, errors will be small if $\vert
Z[f_i-\theta f_f]-Z_{exact} [f_i\theta f_j] \vert$ is small on a
suitable domain of test functions.  One does not have to worry about
errors being amplified by analytic continuation because all
observables are expressed directly in terms of $Z[f_i-\theta f_f]$ or
the corresponding quantities in terms of Euclidean Green functions.
Computational questions are also relevant.  The Euclidean Green
functions require the accurate evaluation of integrals with modest
dimensions, (4(N-1)) where $N$ is the number of particles.  While the
integrands are smooth, the integrals must be calculated accurately,
particularly for scattering calculations.  The dimensions could make a
sufficiently accurate calculation of the matrix elements difficult.
This also needs to be explored using better models.

Another important open question that requires more investigation is
the implementation of this strategy in QCD, where reflection
positivity need not hold on states that are not color singlets.  For
the type of calculations discussed, i.e one particle states and
scattering states, the two-Hilbert space injection operator has range
in the space of physical states and one expects that reflection
positivity holds on this subspace, which is all that is needed to
implement the computational strategies discussed in this paper.  One
interesting observation in QCD, which is often formulated in terms of
Euclidean Green functions, is that there are no problems in dealing
with relative Euclidean times between physical states in the
formulation of scattering theory used in this paper.  This is because
the scattering is actually formulated in Minkowski space, even though
the computations are performed in using Euclidean Green functions.

This work supported in part by the U.S. Department of Energy, under 
contract DE-FG02-86ER40286.


%
%
%
%
%
%

\vfill\eject

\begin{figure}
\begin{center}
\includegraphics[width=1.0\textwidth]{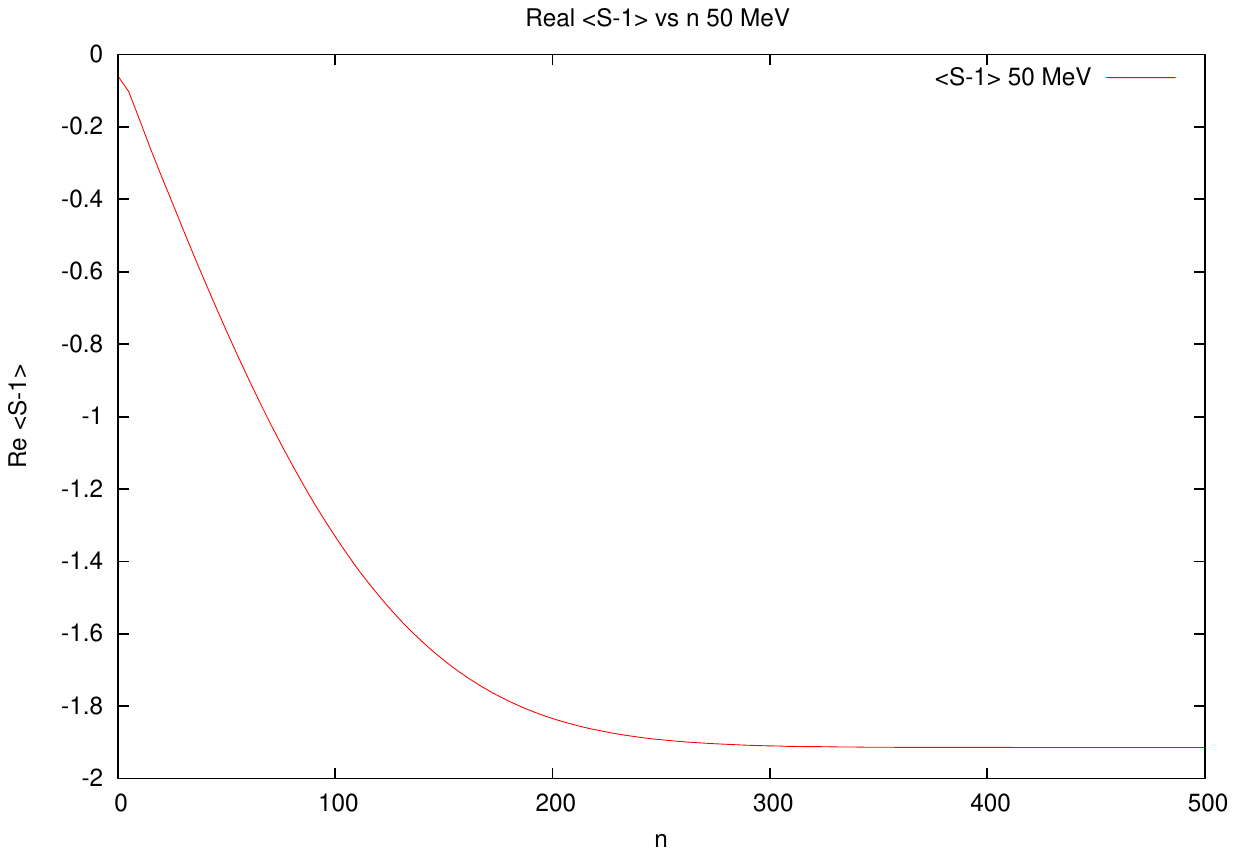}
\caption{Re ($\langle \phi \vert (S-1)\vert \phi \rangle $) vs n 50[MeV] $\alpha=325000$[GeV$^{-2}$]}
\end{center}
\label{fig:1}      
\end{figure}

\begin{figure}
\begin{center}
\includegraphics[width=1.0\textwidth]{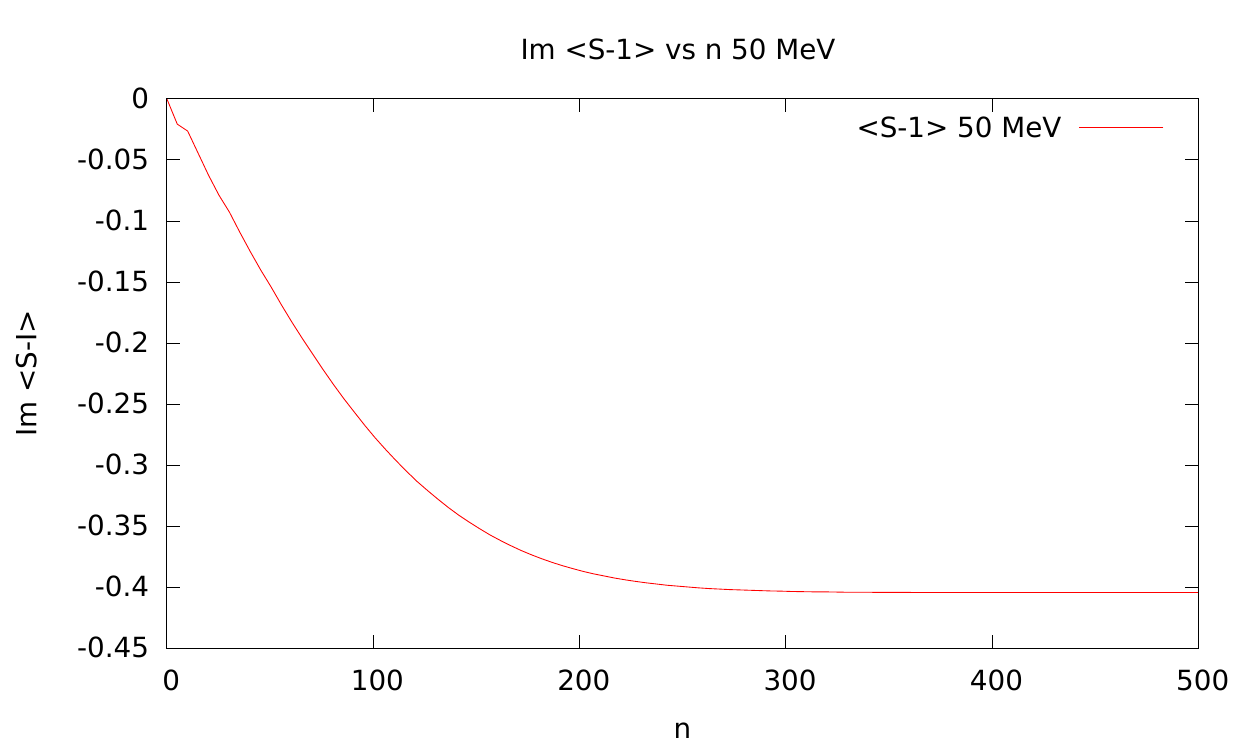}
\caption{Im ($\langle \phi \vert (S-1)\vert \phi \rangle $) vs n 50[MeV] $\alpha=325000$[GeV$^{-2}$]}
\end{center}
\label{fig:2}      
\end{figure}

\begin{figure}
\includegraphics[width=1.0\textwidth]{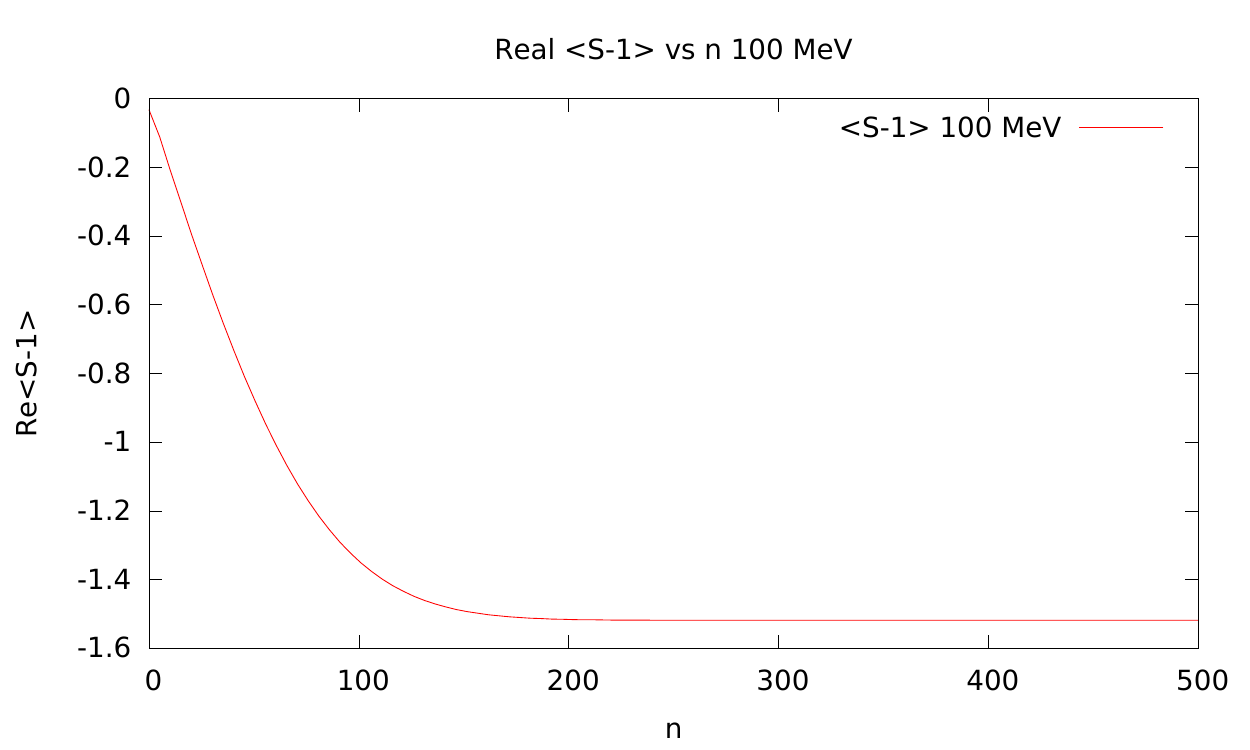}
\caption{Re ($\langle \phi \vert (S-1)\vert \phi \rangle $) vs n 100[MeV] $\alpha=105000$[GeV$^{-2}$]}
\label{fig:3}      
\end{figure}

\begin{figure}
\includegraphics[width=1.0\textwidth]{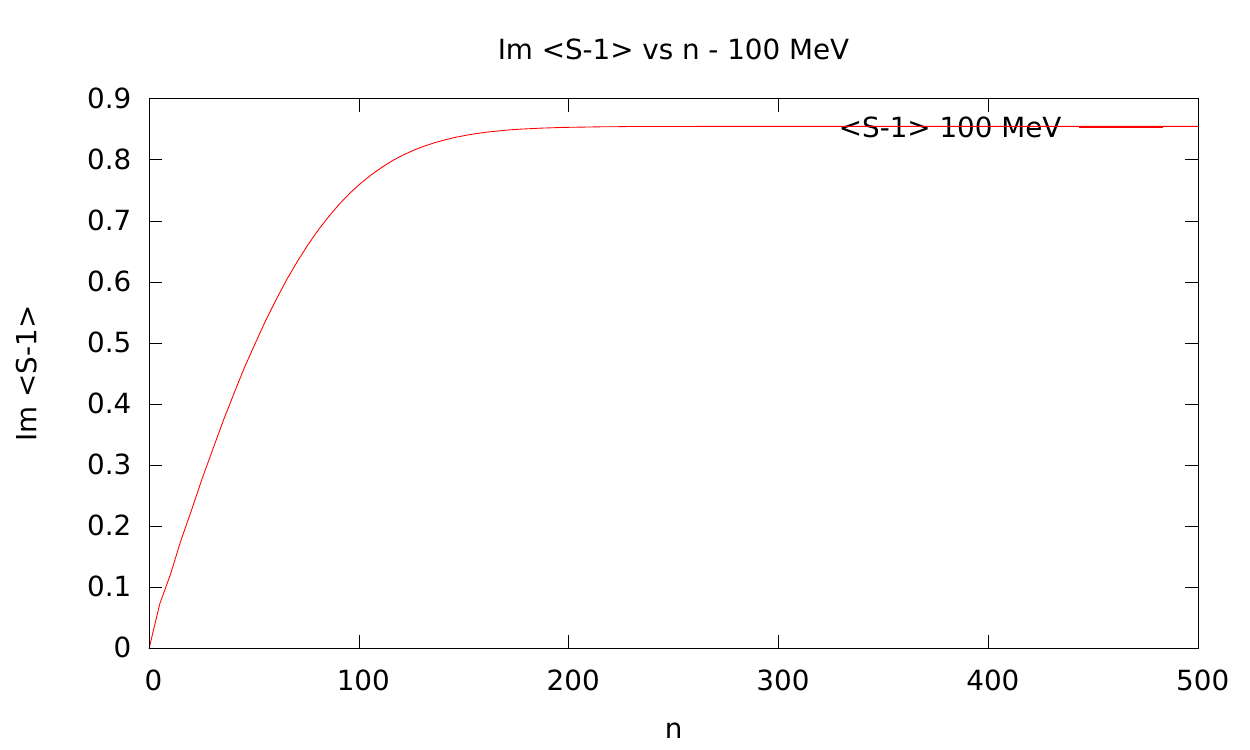}
\caption{Im ($\langle \phi \vert (S-1)\vert \phi \rangle $) vs n 100[MeV] $\alpha=105000$[GeV$^{-2}$]} 
\label{fig:4}      
\end{figure}

\begin{figure}
\includegraphics[width=1.0\textwidth]{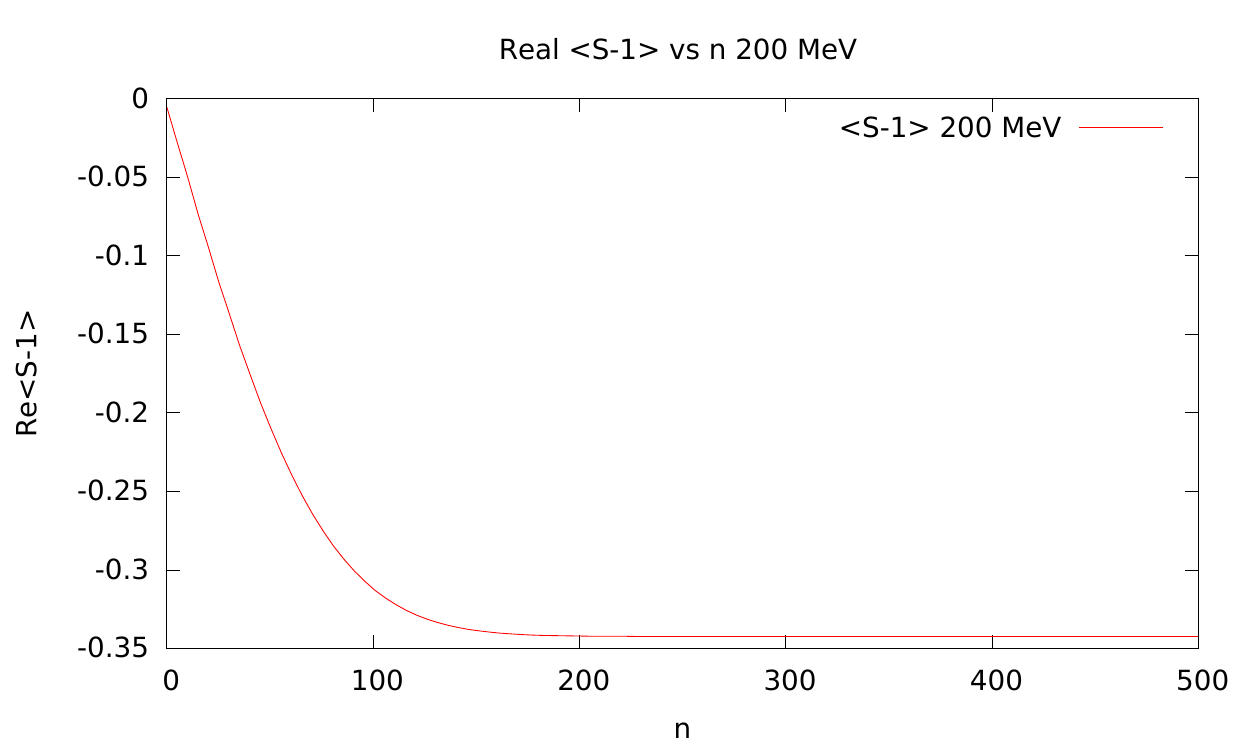}
\caption{Re ($\langle \phi \vert (S-1)\vert \phi \rangle $) vs n 200[MeV] $\alpha=26000$[GeV$^{-2}$]}
\label{fig:5}      
\end{figure}

\begin{figure}
\includegraphics[width=1.0\textwidth]{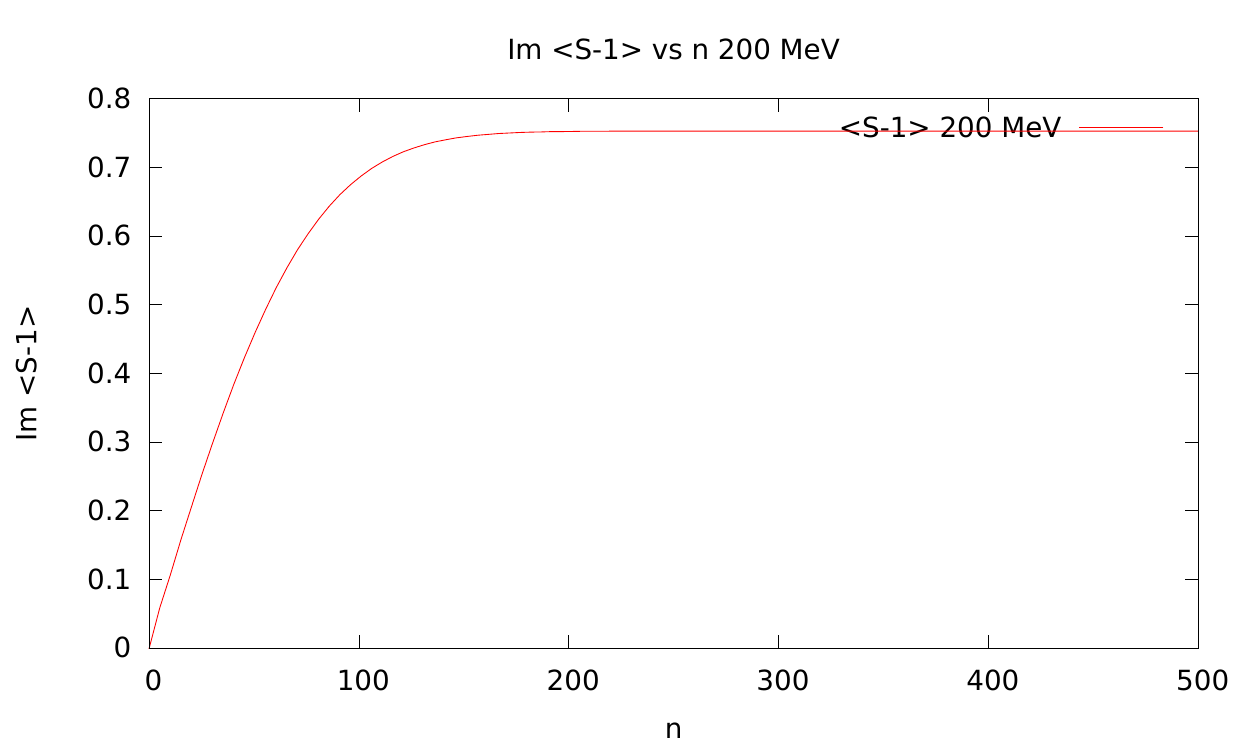}
\caption{Im ($\langle \phi \vert (S-1)\vert \phi \rangle $) vs n 200[MeV] $\alpha=32000$[GeV$^{-2}$]}
\label{fig:6}      
\end{figure}

\begin{figure}
\includegraphics[width=1.0\textwidth]{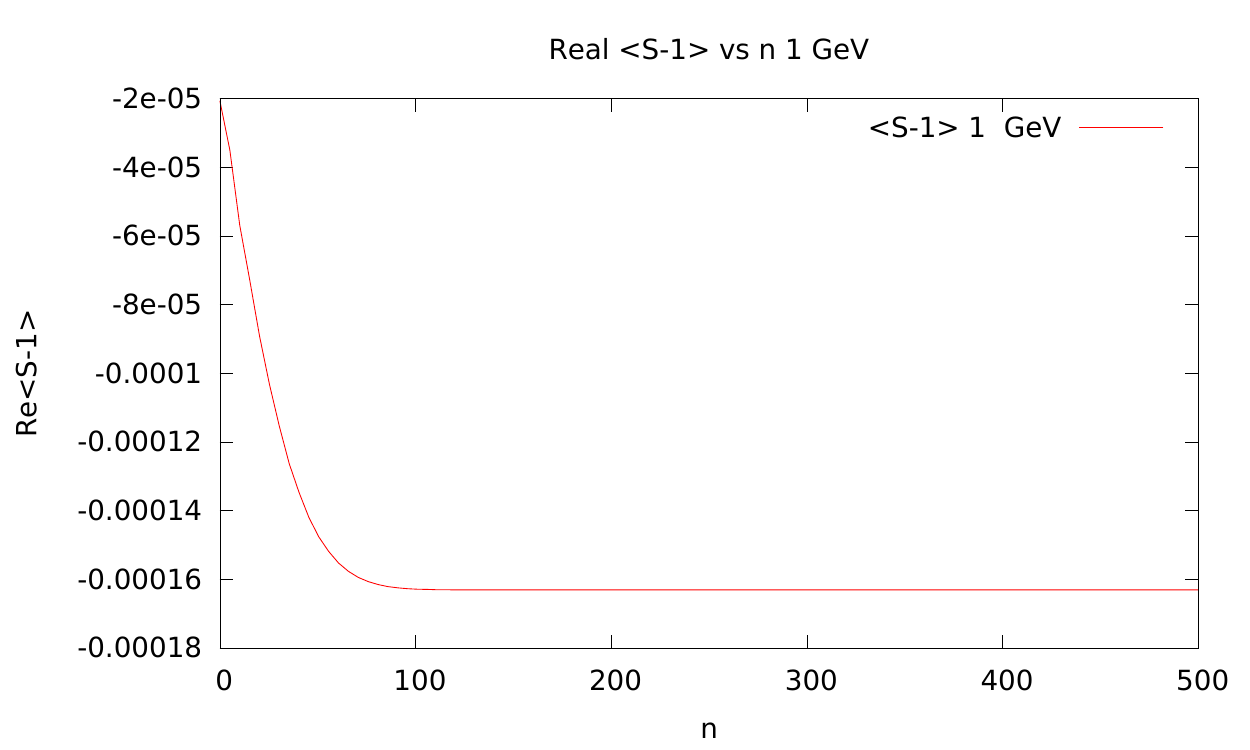}
\caption{Re ($\langle \phi \vert (S-1)\vert \phi \rangle $) vs n 1[GeV] $\alpha=600$[GeV$^{-2}$]}
\label{fig:7}      
\end{figure}

\begin{figure}
\includegraphics[width=1.0\textwidth]{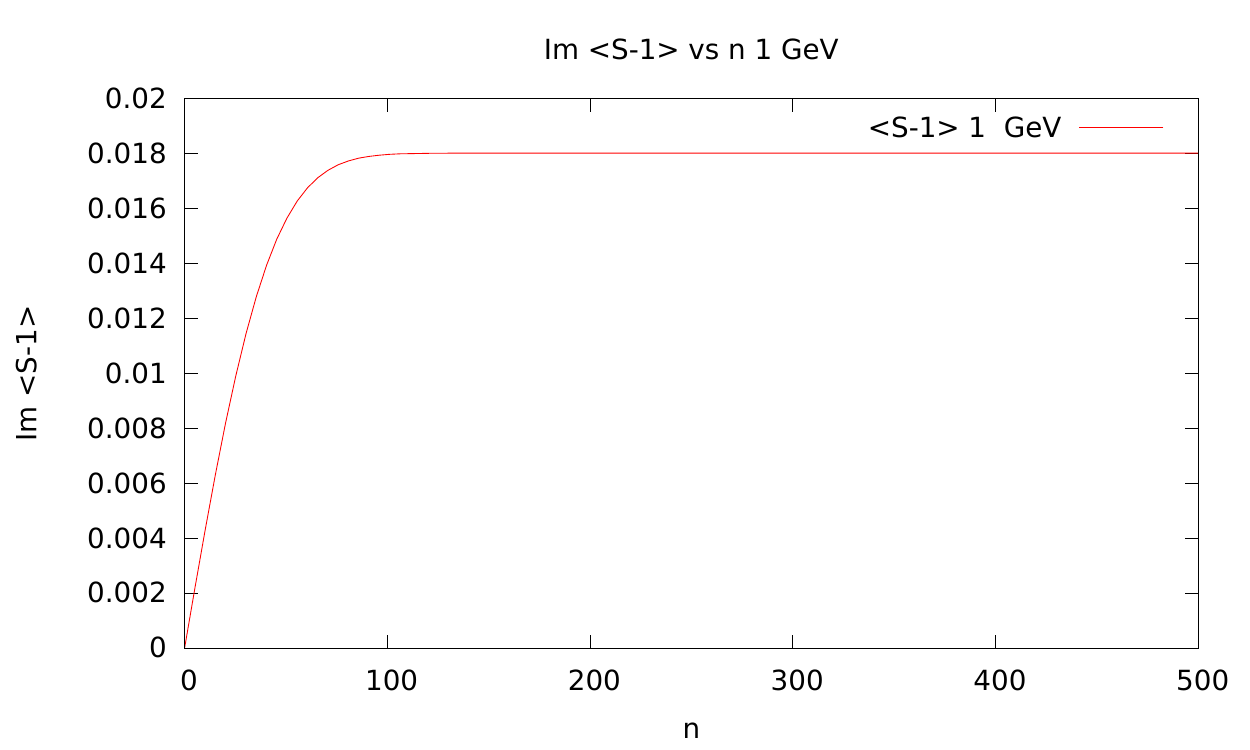}
\caption{Im ($\langle \phi \vert (S-1)\vert \phi \rangle $) vs n 1[GeV] $\alpha=600$[GeV$^{-2}$]}
\label{fig:8}      
\end{figure}

\begin{figure}
\includegraphics[width=1.0\textwidth]{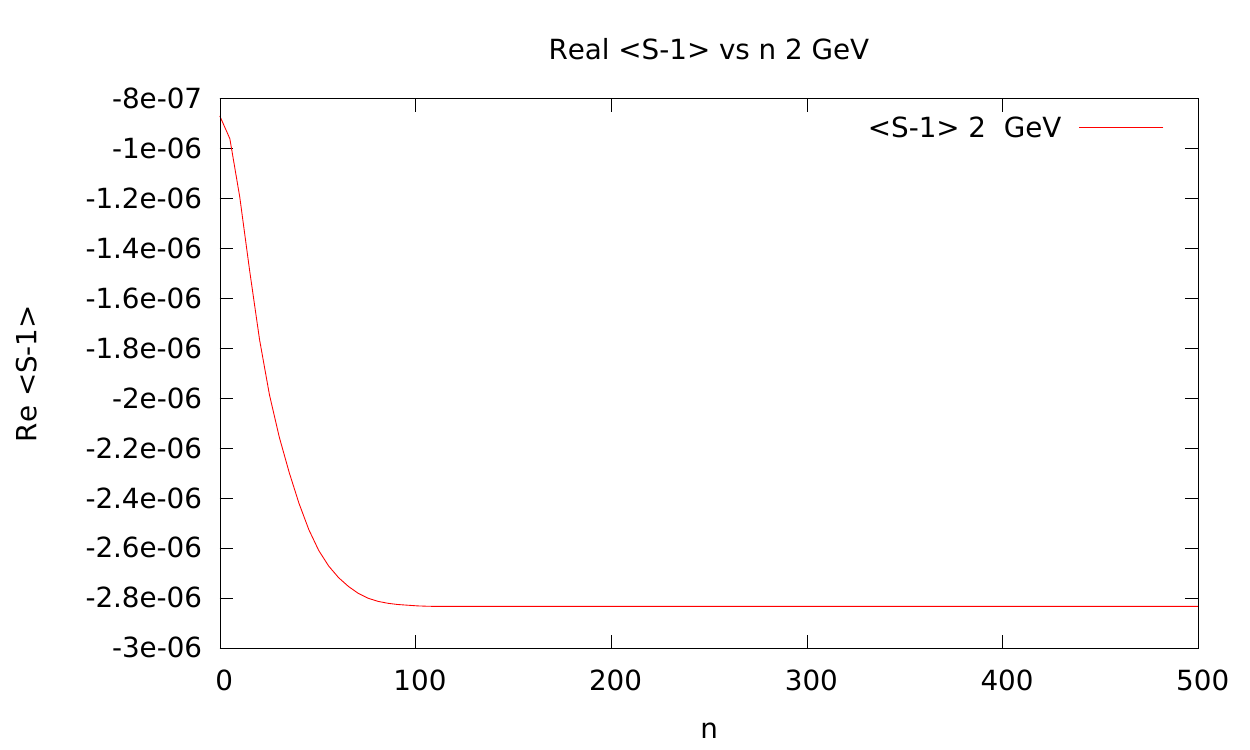}
\caption{Re($\langle \phi \vert (S-1)\vert \phi \rangle $) vs n 2[GeV] $\alpha=250$[GeV$^{-2}$]}
\label{fig:9}      
\end{figure}

\begin{figure}
\includegraphics[width=1.0\textwidth]{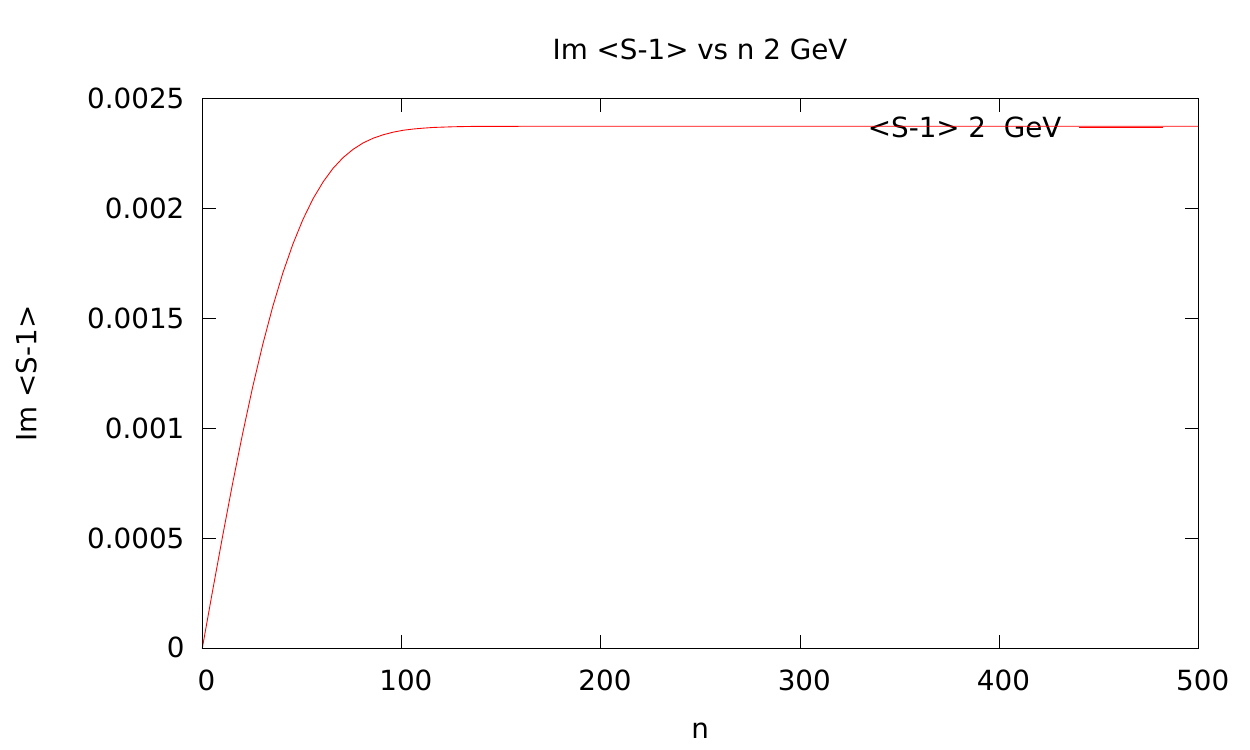}
\caption{Im($\langle \phi \vert (S-1)\vert \phi \rangle $) vs n 2[GeV] $\alpha=250$[GeV$^{-2}$]}
\label{fig:10}      
\end{figure}



\begin{figure}
\begin{center}
\includegraphics[width=1.0\textwidth]{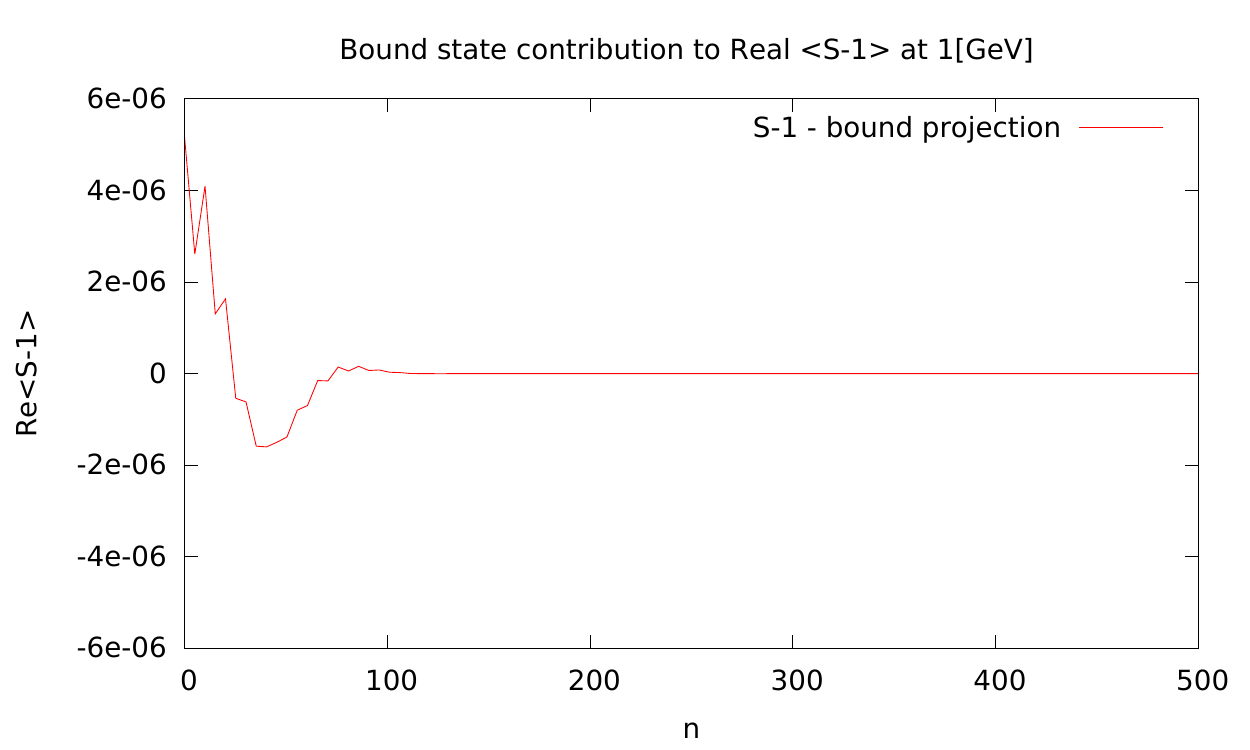}
\caption[Short caption for figure 3]{\label{labelFig11} 
Bound state contribution to Re ($\langle \phi \vert (S-1)\vert \phi \rangle $) vs n 1[GeV] $\alpha=600$[GeV$^{-2}$] 
}
\end{center}
\label{fig.11}
\end{figure}

\begin{figure}
\begin{center}
\includegraphics[width=1.0\textwidth]{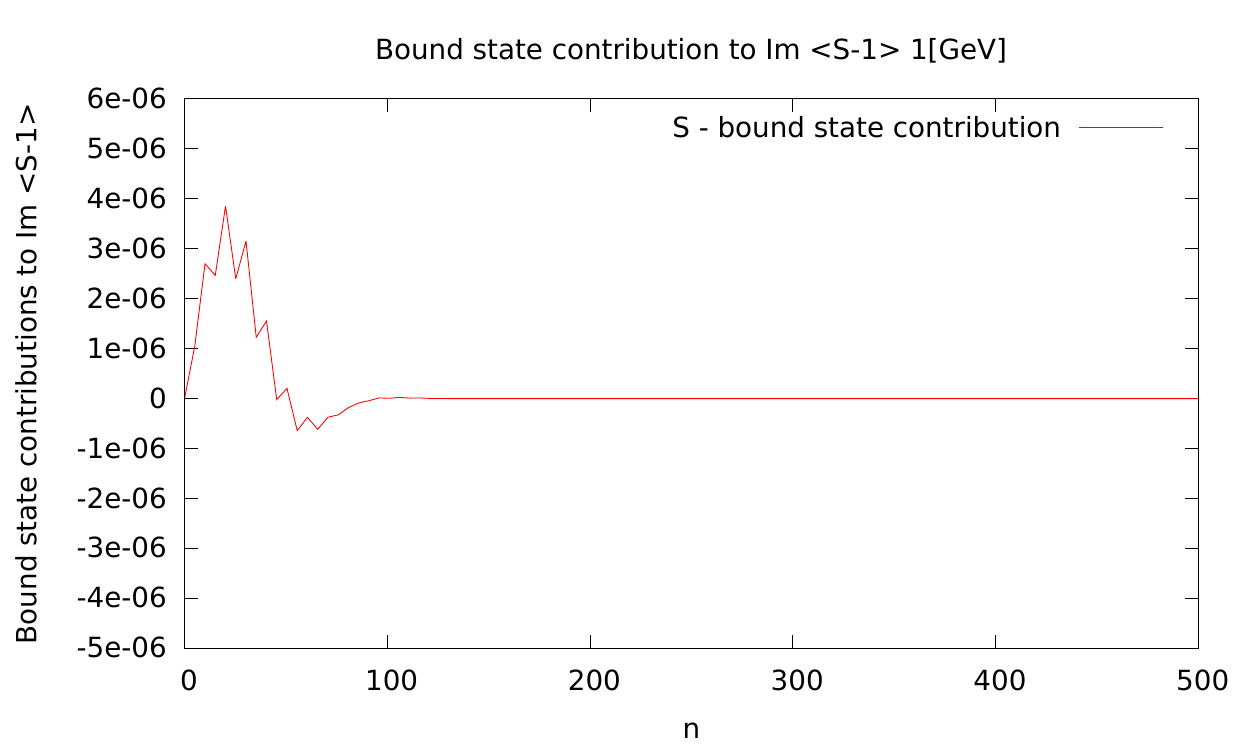}
\caption[Short caption for figure 4]{\label{labelFig12}
Bound state contribution to Im ($\langle \phi \vert (S-1)\vert \phi \rangle $) vs n 1[GeV] $\alpha=600$[GeV$^{-2}$]}
\end{center}
\label{fig.12}
\end{figure}

\end{document}